\documentclass[referee, pdflatex,sn-nature]{sn-jnl}

\usepackage{graphicx}%
\usepackage{multirow}%
\usepackage{amsmath,amssymb,amsfonts}%
\usepackage{amsthm}%
\usepackage{mathrsfs}%
\usepackage[title]{appendix}%
\usepackage{xcolor}%
\usepackage{textcomp}%
\usepackage{manyfoot}%
\usepackage{booktabs}%
\usepackage{algorithm}%
\usepackage{algorithmicx}%
\usepackage{algpseudocode}%
\usepackage{listings}%
\usepackage{soul} 
\usepackage[LGRgreek]{mathastext} 
\makeatletter
\newcommand*{\addFileDependency}[1]{
  \typeout{(#1)}
  \@addtofilelist{#1}
  \IfFileExists{#1}{}{\typeout{No file #1.}}
}
\makeatother

\newcommand*{\myexternaldocument}[1]{%
    \externaldocument{#1}%
    \addFileDependency{#1.tex}%
    \addFileDependency{#1.aux}%
}
\usepackage{xr} 
\myexternaldocument{SI_SF_in_photosynthesis} 

\usepackage{caption}

\usepackage[version=3]{mhchem} 

\begin{document}



\title[Singlet fission contributes to solar energy harvesting in photosynthesis]{\bf Singlet fission contributes to solar energy harvesting in photosynthesis}


\author*[1,2]{\fnm{Shuangqing} \sur{Wang}}\email{shw140@ucsd.edu}
\equalcont{These authors contributed equally to this work.}

\author[3]{\fnm{George A.} \sur{Sutherland}}
\equalcont{These authors contributed equally to this work.}

\author[1]{\fnm{James P.} \sur{Pidgeon}}

\author[3,4]{\fnm{David J. K.} \sur{Swainsbury}}

\author[3]{\fnm{Elizabeth C.} \sur{Martin}}

\author[3]{\fnm{Cvetelin} \sur{Vasilev}}

\author[3]{\fnm{Andrew} \sur{Hitchcock}}

\author[1]{\fnm{Daniel J.} \sur{Gillard}}

\author[1]{\fnm{Ravi Kumar} \sur{Venkatraman}}

\author[1]{\fnm{Dimitri} \sur{Chekulaev}}

\author[1]{\fnm{Alexander I.} \sur{Tartakovskii}}

\author[3]{\fnm{C. Neil} \sur{Hunter}}

\author*[1]{\fnm{Jenny} \sur{Clark}}\email{jenny.clark@sheffield.ac.uk}

\affil[1]{\orgdiv{School of Mathematical and Physical Sciences}, \orgname{University of Sheffield}, \orgaddress{\street{Hicks Building, Hounsfield Road}, \city{Sheffield}, \postcode{S3~7RH}, \country{UK}}}

\affil[2]{\orgdiv{Department of Chemistry and Biochemistry}, \orgname{University of California, San Diego}, \orgaddress{\street{9500 Gilman Drive, MC 0358}, \city{La Jolla, California}, \postcode{92093-0358}, \country{USA}}}

\affil[3]{\orgdiv{School of Biosciences}, \orgname{University of Sheffield}, \orgaddress{\street{Firth Court, Western Bank}, \city{Sheffield}, \postcode{S10~2TN}, \country{UK}}}


\affil[4]{\orgdiv{School of Biological Sciences}, \orgname{University of East Anglia}, \orgaddress{\street{Norwich Research Park}, \city{Norwich}, \postcode{NR4~7TJ}, \country{UK}}}

\maketitle
\textbf{Singlet fission (SF), the spin-allowed conversion of \emph{one} singlet exciton into \emph{two} triplet excitons \cite{Neef2023}, offers a promising strategy for enhancing the efficiency of photovoltaic devices \cite{Nagaya2024,Einzinger2019, Daiber2020, Hanna2006}. However, realising this potential necessitates materials capable of \emph{ultrafast} (sub-picosecond) SF and the generation of \emph{long-lived} ($>$microsecond) triplet excitons \cite{Rao2017}, a synthetic challenge \cite{Pun2019, Korovina2020, Wang2021, Yablon2022, Greissel2023, Orsborne2023, Collins2023}. Some photosynthetic organisms have evolved sophisticated molecular architectures that demonstrate these criteria \cite{Kingma1985,Kingma1985a,Gradinaru2001,Papagiannakis2003,Papagiannakis2002,Akahane2004,Slouf2012a,Niedzwiedzki2016,Yu2017,Niedzwiedzki2017,Niedzwiedzki2020,Zhang2022,Llansola-Portoles2018,Quaranta2021}, but despite 40 years of study, the underlying SF mechanisms and its functional significance in these organisms remain unclear \cite{Yu2017, Gryaznov2019}.
Here, we use a suite of ultrafast and magneto-optical spectroscopic techniques to understand the mechanism of SF within light-harvesting 1 (LH1) complexes from wild-type and genetically modified photosynthetic bacteria. Our findings reveal a SF process, termed ‘heterofission', wherein singlet excitons are transformed into triplet excitons localised on adjacent carotenoid (Crt) and bacteriochlorophyll (BChl) molecules.
We also uncover an unexpected functional role for SF in augmenting Crt-to-BChl photosynthetic energy transfer efficiency. By transiently storing electronic excitation within the SF-generated 
 triplet pair, the system circumvents rapid thermalisation of Crt excitations, thereby enhancing energy transfer efficiency to the BChl Q\textsubscript{y} state, and enabling the organism to usefully harvest more sunlight.} 

\begin{figure}[htbp]
\centering
\includegraphics[width=\textwidth]{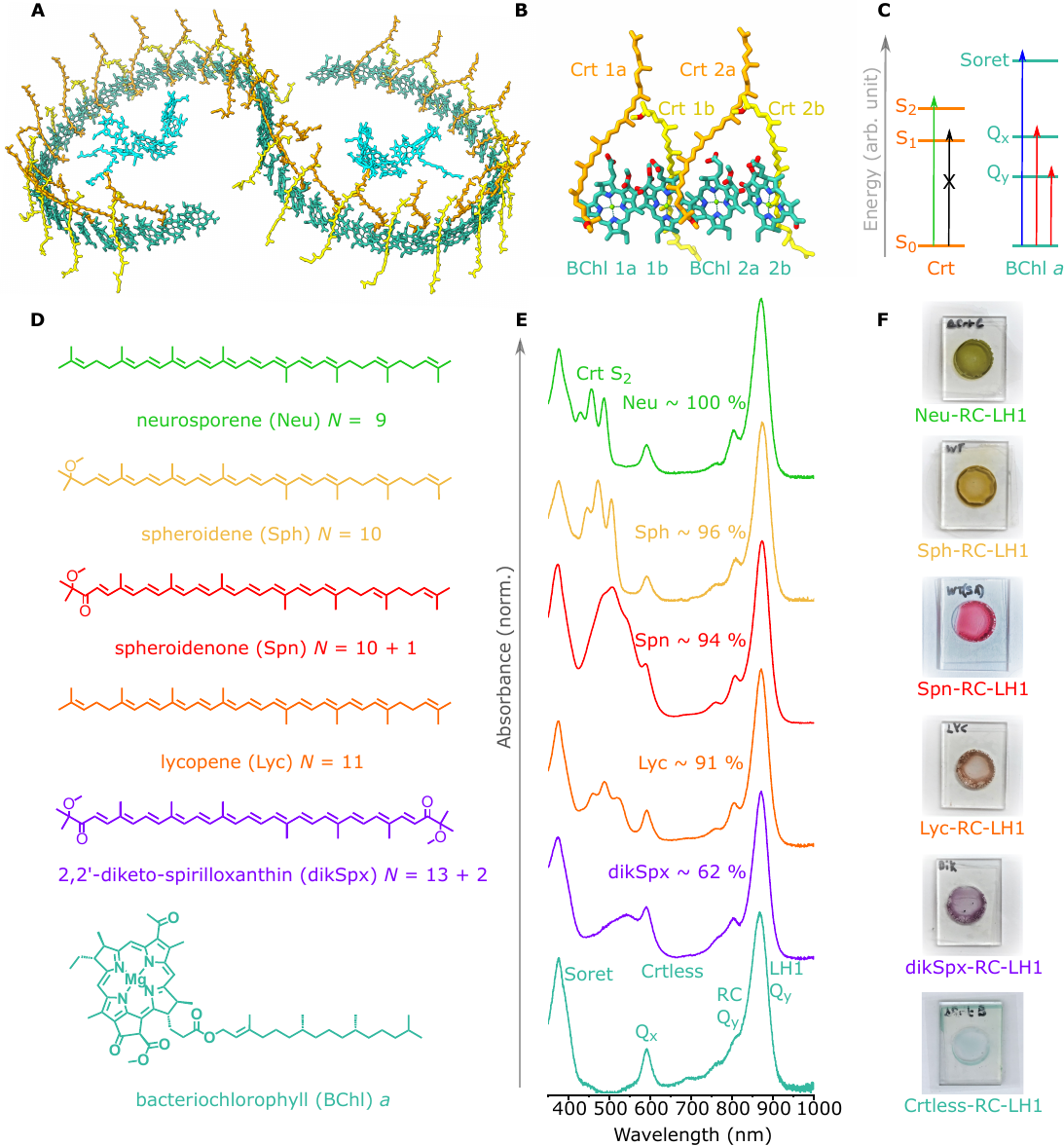}
\caption{\small{\textbf{Reaction-Centre Light-Harvesting 1 (RC--LH1) complexes from \textit{Rba.~sphaeroides}.} (\textbf{A}) Atomic-resolution structure of the dimeric wild-type complex including antenna BChl~\textit{a} (blue-green), Crts (yellow and orange), and RC pigments (cyan), protein residues and BChl~\textit{a} phytol tail removed for clarity (PDBID: 7PQD \cite{Qian2021c}). (\textbf{B}) Arrangement of pigments in an adjacent pair of subunits. Image rendering: ChimeraX \cite{Pettersen2021}. (\textbf{C}) Energy level schematic showing lowest singlet states of Crt (orange) and BChl~\textit{a} (blue-green). Arrows (with cross) denote allowed (forbidden) optical transitions.
(\textbf{D}) Pigment chemical structures, names (acronyms) and conjugation lengths, \textit{N}, (`$+ i$' indicates conjugated \ce{C=O} bonds). (\textbf{E}) Absorption spectra of RC--LH1 complexes (optical path length 10\,mm, OD $\approx$ 0.08 at 872\,nm, Q\textsubscript{y} peak). 
Percentages indicate fraction of Crts in the RC--LH1 complex from the named Crt from Ref.~\cite{Chi2015}. 
(\textbf{F}) Images of RC--LH1 complexes in trehalose/sucrose glass.} \label{fig1}}
\end{figure}

Here we focus on isolated reaction centre light-harvesting 1 (RC--LH1) complexes from the purple photosynthetic bacterium \emph{Rhodobacter (Rba.) sphaeroides}. Purple bacteria are often used as model systems to study fundamental photosynthetic processes, as their adaptable metabolism allows total removal or substitution of photosynthetic pigments and proteins without affecting viability. In addition, the structures of many RC--LH1 complexes have been determined at high resolutions \cite{Qian2021,Qian2021b, Qian2021c, Swainsbury2023} and many mutagenic approaches have been described, including the genetic manipulation of Crts \cite{Chi2015, Sutherland2022}. RC--LH1 complexes from \textit{Rba.~sphaeroides} are known to demonstrate SF \cite{Yu2017} and the wild-type pigment-protein chromophore structure, including the Crts, has been recently published \cite{Qian2021}, reproduced in Figs.~\ref{fig1}~A,B. 

The RC--LH1 complexes shown in Fig.~\ref{fig1} were isolated, as described previously  \cite{Sutherland2022}, from wild-type cells grown under semi-aerobic heterotrophic conditions, in which spheroidenone (Spn) is the predominant carotenoid (Crt) produced, and from engineered strains that produce complexes containing predominantly neurosporene (Neu), spheroidene (Sph), lycopene (Lyc) or 2,2'-diketo-spirilloxanthin (dikSpx) \cite{Chi2015}. We also purified complexes from a 15-\textit{cis}-phytoene synthase mutant ($\Delta$\emph{crtB}), yielding carotenoidless (Crtless) RC--LH1 complexes \cite{Chi2015}.

 The main pigments in these complexes are Crts and BChl~\textit{a}, whose low-energy singlet electronic states are shown in Fig.~\ref{fig1}~C, with chemical structures depicted in Fig.~\ref{fig1}~D. The absorption spectra of the RC--LH1 complexes, Fig.~\ref{fig1}~E, show Crt S\textsubscript{2} absorption ($450-550$\,nm) that red-shifts as the Crt conjugation length increases and three BChl~\textit{a} absorption bands. The BChl~\textit{a} Soret band absorbs around 390\,nm, Q\textsubscript{x} at $\sim600$\,nm and Q\textsubscript{y} around $800-900$\,nm. The Q\textsubscript{y} absorption feature is sensitive to structural changes \cite{Cogdell2002}, thus its similarity in the spectra in Fig.~\ref{fig1}~E suggests the structure of the RC--LH1 complex remains the same, independent of the Crt species incorporated. The Crts show similar resonance Raman spectra in the `conformational fingerprint' region (Fig.~S1) \cite{Yu2017}, suggesting that the Crts within the complexes adopt similar conformations. We therefore make the assumption that the conformations of the proteins and the pigments within them are broadly independent of Crt.

To protect Crtless--RC--LH1 and other complexes from photobleaching under laser illumination, samples were encapsulated in trehalose/sucrose glass, as described previously \cite{Sutherland2020}, for all spectroscopic techniques unless stated otherwise. Photographs of the samples are shown in Fig.~\ref{fig1}~F. Trehalose maintains native protein structure in dehydrated conditions, giving equivalent spectroscopic observations compared with aqueous solutions (Fig.~S2) \cite{Kurashov2018,Jain2009}.

\subsection*{Singlet fission yield decreases as carotenoid conjugation length increases}

\begin{figure}[!h]%
\centering
\includegraphics[width=\textwidth]{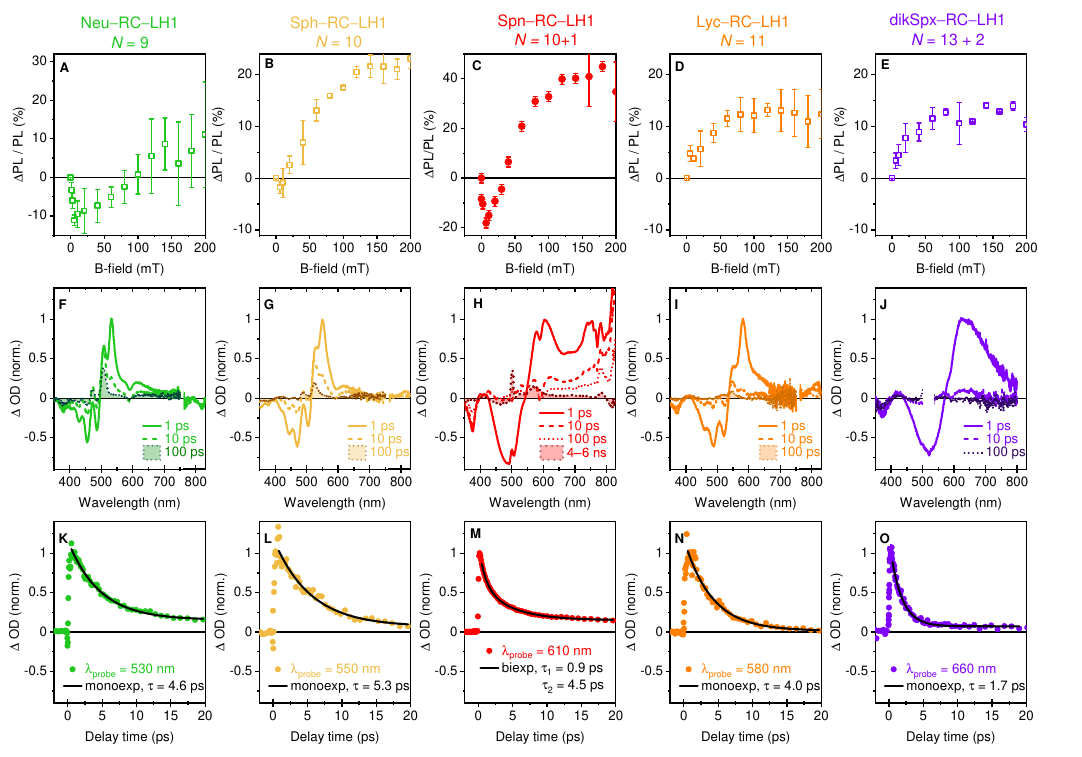}
\caption{\small{\textbf{Spectroscopy of Crt--RC--LH1 complexes as a function of Crt conjugation length.} (\textbf{A--E}) Magnetic field effect (MFE), measured as the relative change in PL intensity, normalised to zero-field for the Crt--RC--LH1 complexes marked in the titles. In panels \textbf{A--C} the negative feature in the MFE demonstrates the presence of SF, in \textbf{D, E} the signal is always positive and is instead dominated by radical pair behaviour, see text for details. Detection at $880$\,nm, $20-40$\,ns delay, excitation at $\lambda_{\mathrm{ex}}=490$\,nm for Neu, $\lambda_{\mathrm{ex}}=520$\,nm for others. Error bars: $2\sigma$ (standard deviation) from subsequent field up and down scans. (\textbf{F--J}) Transient absorption spectra of Crt--RC--LH1 complexes at $\Delta\mathrm{t}=$~$1, 10, 100$\,ps and, for Spn--RC--LH1, $\Delta\mathrm{t}$ averaged over $4-6$\,ns. The shaded spectra show triplets assigned to SF; the relative yield decreases with increasing conjugation length. $\lambda_{\mathrm{ex}}=500$\,nm for Neu--RC--LH1 and  $\lambda_{\mathrm{ex}}=520$\,nm for others. (\textbf{K--O}) Transient absorption dynamics at the peak of the S\textsubscript{1}$\rightarrow$S\textsubscript{n} excited state absorption (ESA) band (markers). Solid lines are mono-exponential fits with lifetimes marked in the legend. Excitation conditions as in (F--J). All complexes were measured following encapsulation in trehalose glass.}}\label{fig2}
\end{figure}

We begin by characterising SF in these complexes using magnetic field-dependent experiments similar to those used to identify SF in \textit{Rba.~sphaeroides} RC--LH1 complexes in the 1980s \cite{Kingma1985,Kingma1985a}. Our magnetic-field dependent measurements for different RC--LH1 complexes with Crt excitation are shown in Figs.~\ref{fig2}~(A--E). The normalised change in photoluminescence (PL) intensity, detected at 880\,nm and $20-40$\,ns after excitation, is plotted as a function of applied magnetic field. The results broadly reproduce the initial measurements \cite{Kingma1985,Kingma1985a}, but extend earlier studies to a wider range of Crt conjugation lengths, and at a specific delay time after excitation.

The magnetic field effect (MFE) shape of the RC--LH1 complex in Figs.~\ref{fig2}~(A--E) varies with Crt conjugation length. The complexes containing the three shortest conjugation length Crts (Neu, Sph, Spn) show a negative feature (reduction in PL) at low magnetic fields, followed by an increase in PL at higher fields, saturating above 125\,mT. This MFE shape is indicative of SF, as originally explained by Merrifield and Johnson \cite{Merrifield1971,Johnson1970} and used by Kingma \emph{et al.,} to assign SF in RC--LH1 complexes \cite{Kingma1985, Kingma1985a}. The detailed differences in the shape of the SF-associated MFEs are due to differences in the Crt zero field splitting parameters, the rates of SF and its reverse, triplet-triplet annihilation \cite{Bossanyi2022}, and the relative contribution to the MFE from SF and radical pair recombination (see below). Quantification of these effects is beyond the scope of the present work. 

On the other hand, the RC--LH1 complexes containing longer conjugation length Crts (Lyc and dikSpx) show a monotonic increase in PL with applied field, saturating above 50\,mT, without displaying negative values at any applied field. 
This latter behaviour is due to radical pairs (RP) within the RC \cite{Eichwald1996,Steiner1989,Hoff1981,Hoang2018} and indicates that little or no SF occurs in these complexes. 

Recent work on SF has highlighted the importance of exchange coupling between the SF-generated triplets on experimental observables such as the MFE \cite{Bayliss2016, Bayliss2018, Ishikawa2018, Bossanyi2022}. Indeed, the presence of an SF-induced MFE at fields below 200\,mT, as observed here, indicates that the SF-generated triplets are weakly exchange coupled  \cite{Benk1981, Musser2019} and demonstrate inter-triplet exchange interactions (\textit{J}) that are much smaller than the intra-triplet dipolar zero-field splitting parameter (\textit{D}, $\sim 10\,\mu$eV) \cite{Benk1981,Musser2019,Bossanyi2021b,Gryaznov2019}. As \textit{J} depends on orbital overlap \cite{Benk1981, Kollmar1993, Blundell2021}, the presence of this low-field MFE demonstrates that the SF-generated triplets  have negligible orbital overlap \cite{Bayliss2018,Bossanyi2021b} and thus \emph{do not} reside on the same chromophore 
\cite{Sutherland2023, Gryaznov2019}, counter to the currently accepted wisdom \cite{Gradinaru2001,Papagiannakis2002,Yu2017,Niedzwiedzki2017,Zhang2022}. 

The fact that SF does not occur along a single Crt molecule, twisted or otherwise \cite{Sutherland2023, Gryaznov2019}, leaves open the question of the mechanism of SF in these light-harvesting complexes. Having ruled out SF occurring along a single Crt chain, the next two most obvious mechanisms are that SF occurs between two neighbouring Crts, or SF occurs between neighbouring Crt and BChl~\textit{a} molecules. The first option - intermolecular SF between Crts - can be ruled out because studies indicate that SF occurs even in RC--LH1 complexes from \emph{Rsp.~rubrum} \cite{Rademaker1980, Kingma1985, Papagiannakis2003}, whose Crts are separated by $>10\,\AA$ \cite{Qian2021a}, a distance at which intermolecular SF is extremely unlikely. The second option -- hetero-SF, where a single Crt excited state produces a pair of triplets residing on neighbouring Crt and BChl~\textit{a} molecules -- was initially ruled out by transient absorption (TA) spectroscopy studies that showed no evidence of BChl~\textit{a} triplet features following SF \cite{Gradinaru2001}. 

Figs.~\ref{fig2}~(F--O) shows a selection of Crt--RC--LH1 transient absorption spectra and dynamics following Crt excitation (pump wavelengths marked in the caption). 
By 1\,ps after excitation, several absorption and ground state bleach (GSB) features are apparent. The dominant feature in all complexes except Spn--RC--LH1 is the Crt S\textsubscript{1}$\rightarrow$S\textsubscript{n} excited state absorption (ESA) and associated GSB, with peak ESA at 530\,nm (Neu), 550\,nm (Sph), 610\,nm (Spn), 580\,nm (Lyc), and 660\,nm (dikSpx). Spn--RC--LH1 additionally shows evidence of an intramolecular charge transfer (ICT) state, with absorption bands in the near-infrared (NIR) \cite{Slouf2012a} \footnote{As an aside, we note that there is significant configuration interaction (mixing) between electronic states in Crts due to the strong electronic and vibronic interactions \cite{Barford2013, Valentine2020, Barford2022, Manawadu2022}. Here, for simplicity, we give specific names to diabatic states (e.g.~S\textsubscript{1}, ICT,\ldots), but implicitly assume interaction between them throughout the paper (forming adiabatic mixed states).}. 

These Crt S\textsubscript{1} and ICT states decay on a picosecond timescale, as shown in Figs.~\ref{fig2}~(K--O). In these complexes, the S\textsubscript{1} lifetime lies between $4-5.5$\,ps for Crts with conjugation lengths $\textit{N}=9-11$, reflecting two competing decay processes: non-radiative decay to the ground state \cite{Polli2004}, and S\textsubscript{1}$\rightarrow$Q\textsubscript{y} energy transfer \cite{Cong2008}, which we discuss further below. 

With the exception of dikSpx--RC--LH1, where we observe no Crt triplets on these timescales, by 1\,ps all other Crt--RC--LH1 spectra also show evidence of Crt T\textsubscript{1}$\rightarrow$T\textsubscript{n} excited state absorption and associated GSB. These features dominate the spectra after the Crt singlet states (S\textsubscript{1} and ICT) have decayed, highlighted by the shaded spectra in Fig.~\ref{fig2} \footnote{T\textsubscript{1}$\rightarrow$T\textsubscript{n} spectra are assigned by comparison with sensitised triplet spectra in Fig.~S11 with peaks at 505\,nm (Neu), 525\,nm (Sph), 570\,nm (Spn), 545\,nm (Lyc), absent in dikSpx}. The sub-ps formation of Crt triplets suggests they are generated by SF \cite{ Wang2010,Wang2011,Musser2015,Llansola-Portoles2018,Quaranta2021,Gradinaru2001,Papagiannakis2003,Papagiannakis2002,Akahane2004,Slouf2012a,Niedzwiedzki2016,Yu2017,Niedzwiedzki2017,Niedzwiedzki2020,Zhang2022}, confirming our MFE results. We find that shorter conjugation length Crts show higher triplet yield, in contrast with earlier reports \cite{Akahane2004}. These yields are quantified below, see Supplementary Section~7.2 for details.

All the TA spectra in Figs.~\ref{fig2}~(F--J) also show evidence of BChl~\textit{a} bleach (Soret 378\,nm, Q\textsubscript{x} 590\,nm, Q\textsubscript{y} $750-850$\,nm) and excited state absorption signatures (across the visible range), indicating that upon Crt excitation, energy transfers to BChl~\textit{a} molecules within 1\,ps.  These BChl~\textit{a} features are relatively weak, mainly due to the $\sim10\times$ lower excited state extinction coefficient of BChl~\textit{a} compared to Crts in the visible spectral region, see Fig.~S3 \cite{Cogdell1983, Borland1987}. Interestingly, the BChl~\textit{a} spectral features between $750-830$\,nm change as a function of the type of Crt: from negative (GSB) in Neu--RC--LH1 to positive (ESA) in Lyc--RC--LH1. We show below that this change in spectra reflects the Crt-conjugation-length dependence of prompt BChl~\textit{a} triplet formation, indicative of the presence of heterofission.

\newpage
\subsection*{Hetero singlet fission generates triplets on neighbouring carotenoid and bacteriochlorophyll~\textit{a} molecules}

\begin{figure}[htpb]%
\centering
\includegraphics[width=0.5\textwidth]{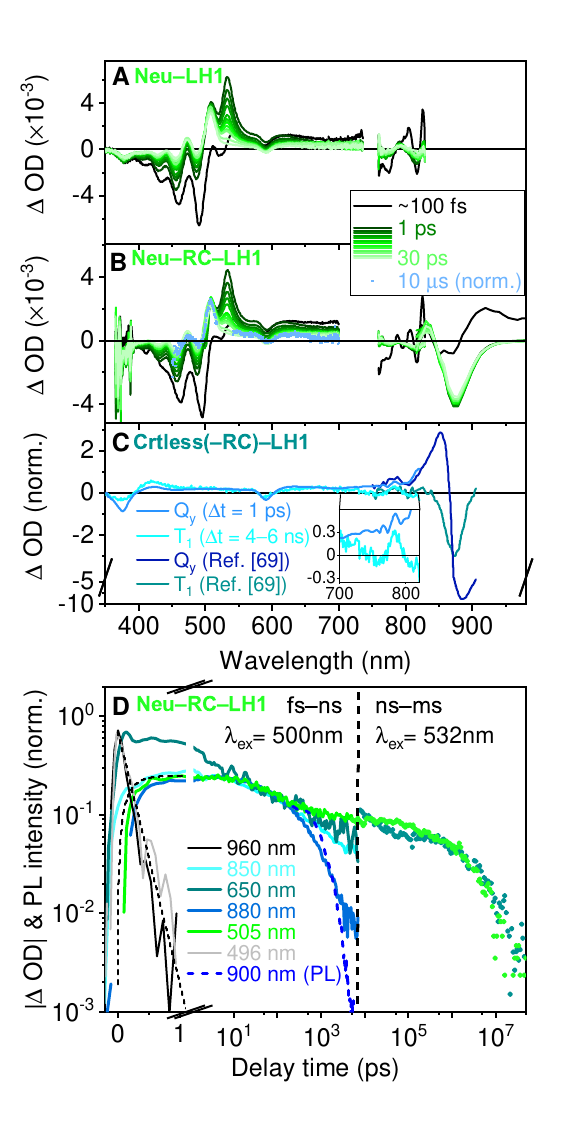}
\caption{\textbf{Transient absorption (TA) spectroscopy of Neu and Crtless (RC-)LH1 complexes.} \textbf{(A)} TA spectra of Neu--LH1, $\lambda_{\mathrm{ex}}=500$\,nm measured with two probe windows (UV--Vis, $750-830$\,nm). Green spectra taken at $\Delta t=1,2,3,4,5,6,8,9,10,12,20,30$\,ps delay time. \textbf{(B)} TA spectra of Neu--RC--LH1 over three probe windows (UV--Vis, $750-830$\,nm and $800-1300$\,nm), $\Delta t=10$\,$\mu$s spectrum normalised to the 30\,ps spectrum (blue markers, $\lambda_{\mathrm{ex}}=532$\,nm for the $\mu$s data), other details as (A). \textbf{(C)} BChl~\textit{a} TA spectra associated with Q\textsubscript{y} and T\textsubscript{1} population in a Crtless--RC--LH1 measured in this work ($\lambda_{\mathrm{ex}}=590$\,nm) and Crtless--LH1 reproduced from Ref.~\cite{Uragami2020}. Note the break in the y-axis scale. Inset shows zoom-in of fingerprint region: T\textsubscript{1} has negative features (GSB) while Q\textsubscript{y} is positive (ESA). Analysis and assignments in Supplementary Section~4. \textbf{(D)} TA and photoluminescence (PL) transients of Neu--RC--LH1 complexes. Lines: fs--ns transients normalised at 700\,ps (except 960\,nm and 496\,nm). Markers: ns--ms TA data normalised at 1\,$\mu$s. Black dashed lines: monoexponential decay and rise, both with $\tau=170$\,fs. Detection wavelengths marked in the legend.
}\label{fig3}
\end{figure}

To better assign the weak BChl~\textit{a} transient absorption features to singlet, triplet or other species, we compare transient absorption spectra of a Crtless mutant (Fig.~\ref{fig3}~C) to more detailed TA spectra of the highest-SF-yield complex, Neu(--RC)--LH1 (Figs.~\ref{fig3}~A,B).

The TA spectra of Neu--LH1 with and without RC in panels B and A, respectively, are similar, as expected. In both cases, the earliest recorded spectra (black lines) show signatures of S\textsubscript{2} population (stimulated emission $\sim496$\,nm and excited state absorption in the NIR) that rapidly decay leaving Neu S\textsubscript{1}, Neu T\textsubscript{1} and BChl~\textit{a} populations, as previously reported  \cite{Niedzwiedzki2016}. Interestingly, the spectral shape for Neu--RC--LH1 at 30\,ps in the visible matches that measured at 10\,$\mu$s (markers, a delay time usually associated with triplet population) indicating very little spectral evolution beyond 30\,ps and suggesting the presence of BChl~\textit{a} triplets on picosecond timescales. (See Fig.~S13~A for the full ns--ms dataset).

 For comparison, panel~C shows the transient absorption spectra associated with BChl~\textit{a} Q\textsubscript{y} (blue) and T\textsubscript{1} (teal) population, measured in Crtless--RC--LH1 complexes in this work and in Crtless--LH1 complexes in Ref.~\cite{Uragami2020} (spectra are normalised in the overlap region between $750-800$\,nm). Detailed analysis of the spectra is available in Supplementary Section~4.2 and Ref.~\cite{Uragami2020}.

The spectra associated with Q\textsubscript{y} and T\textsubscript{1} population are strikingly similar in the visible spectral region, as is well known for porphyrins and chlorins \cite{Kosumi2018}, due to their symmetry \cite{Gouterman1960, Tobita1984} (it is not an artefact from two-photon or sequential excitation effects, although these are apparent, see Fig.~S9). The similarity makes it difficult to assign BChl~\textit{a} spectral signatures in Crt--RC--LH1 complexes to either the singlet (Q\textsubscript{y}) or triplet (T\textsubscript{1}) population. We therefore focus on the NIR spectral range where the spectra diverge. Between $700-830$\,nm, for example (see inset), population of the Q\textsubscript{y} state results in an absorption (positive), while population of T\textsubscript{1} results in weak bleach features (negative), making this a useful fingerprint region for distinguishing the Q\textsubscript{y} and T\textsubscript{1} populations. Likewise, the dominant Q\textsubscript{y} bleach peak ($830-950$\,nm) is red-shifted and over 2.5$\times$ stronger when  Q\textsubscript{y} is populated compared to when T\textsubscript{1} is populated (note the break in the y scale bar in Fig.~\ref{fig3}~C). 

Comparing Figs. \ref{fig3}~A,B and \ref{fig3}~C, shows that the BChl~\textit{a} components of the early-time Neu(--RC)--LH1 spectra in panels A/B are most similar to the T\textsubscript{1} reference spectrum in panel C, exhibiting negative bleach features between $750-830$\,nm and a dominant GSB peak at 870\,nm (for completeness we have plotted a direct comparison in Fig.~S8). This suggests that, upon Crt S\textsubscript{2} excitation in Neu--RC--LH1, BChl~\textit{a} triplets are formed within a picosecond, i.e.~hetero-SF is occurring in Crt--RC--LH1 complexes, generating triplets on neighbouring Crt and BChl~\textit{a} molecules.

The dynamics of the transient absorption features in Neu--RC--LH1 are shown in Fig.~\ref{fig3}~D (TA), compared to ps--ns photoluminescence spectroscopy (PL, dashed) at the detection wavelengths marked in the legend. We use single wavelengths here as global analysis showed significant correlations between species (see Supplementary Section~5.1 for details). In the ps--ns regime, these single wavelength transients 
are normalised at 700\,ps (except 960\,nm and 496\,nm which have decayed by then), while the ns--ms transients are normalised at 1\,$\mu$\,s, to directly compare both the early-time dynamics (ps--ns) and those of the final decaying state (ns--ms).

The time-resolved spectral decays in the early-time range (ps--ns), to the left of the vertical dashed line in Fig.~\ref{fig3}~D, show that the Neu S\textsubscript{2} population (excited state absorption at 960\,nm and stimulated emission at 496\,nm) decays monoexponentially with a $\sim170$\,fs time constant (dashed line fit). 
Except 650\,nm, where S\textsubscript{2} also absorbs \cite{Polli2004}, the other TA transients rise on a similar timescale ($\tau=170$\,fs monoexponential rise shown by black dashed line),  suggesting direct population of all other states from S\textsubscript{2}. 

The dynamics at 850\,nm (cyan), within the Q\textsubscript{y} GSB band, are similar to the Neu T\textsubscript{1} population dynamics measured at 505\,nm (green). The BChl~\textit{a} excited state absorption band at 650\,nm (teal), where Neu T\textsubscript{1} should not absorb \cite{Niedzwiedzki2020}, also shows similar decay dynamics. These long-lived dynamics confirm our earlier assignment of these features being dominated by BChl~\textit{a} triplets, with minor contribution from the Q\textsubscript{y} population.

This assignment is further supported by the slow $\mu$s decay of the 650\,nm absorption band (teal) shown to the right of the dashed line, which decays in tandem with Neu T\textsubscript{1} population (505\,nm, green) from $15\,\mathrm{ns}-50\,\mu$s, indicating that BChl~\textit{a} triplets decay together with Neu triplets over several orders of magnitude in time. \footnote{The 505\,nm and 650\,nm transients behave differently between $1-15$\,ns because the long-time (ns--ms) and short-time (ps--ns) dynamics were measured with different excitation conditions, artefacts from which occur in this temporal overlap region (see Supplementary Section~5.3 for details).} The dynamics therefore show that between $<1\,\mathrm{ps}-10$\,ns and $15\,\mathrm{ns}-50\,\mu$s timescales, both BChl~\textit{a} and Neu triplet states decay in tandem. 

The triplet lifetimes we observe here (half life of $\sim0.3-0.4$\,$\mu$s) are shorter than those typically observed for isolated Neu or BChl~\textit{a} triplets \cite{Kim2007}. We attribute this relatively short lifetime to the presence of \emph{triplet pairs}, which have shorter lifetimes than isolated triplets as they can decay in tandem \cite{Musser2015}.  This SF-associated shortening of the lifetime is supported by the observation in Fig.~S13 that triplet lifetimes in RC--LH1 complexes which demonstrate SF (Neu- and Sph--RC--LH1) are almost an order of magnitude shorter than triplet lifetimes measured in complexes which demonstrate negligible SF (Lyc- and dikSpx--RC--LH1). These latter complexes exhibit Crt triplet population following triplet exciton transfer from BChl~\textit{a} in the conventional way \cite{Kakitani2007, Slouf2012a}.

We end our discussion of Fig.~\ref{fig3} by noting that the TA dynamics measured at 880\,nm, near the peak of the GSB associated with Q\textsubscript{y} population, decay similarly to the Q\textsubscript{y} PL (dashed line), with only a minor long-lived component, indicating that both transients are dominated by Q\textsubscript{y} population decay. The presence of some Q\textsubscript{y} population on picosecond timescales shows that Q\textsubscript{y} states are also populated from Crt excitation \emph{via} either direct singlet exciton energy transfer, as described previously \cite{Niedzwiedzki2017, Cong2008}, or rapid transfer from the triplet pair states to Q\textsubscript{y}. As population of the Q\textsubscript{y} state leads to a strong GSB feature (more than $2.5\times$ stronger than with BChl triplet population), any Q\textsubscript{y} population is likely to dominate BChl~\textit{a} GSB signatures in the NIR, making it difficult to observe the weaker triplet population signatures \cite{Gradinaru2001}.

In summary, comparison between transient absorption spectroscopy of Neu(--RC)--LH1 and Crtless--LH1 complexes suggests that within a couple of hundred fs of exciting the Neu S\textsubscript{2} state,  BChl~\textit{a} and Neu triplets are formed and decay in tandem. Q\textsubscript{y} states are also formed, \emph{via} either rapid conversion from the triplet pair states or direct singlet exciton energy transfer from the Neu S\textsubscript{1} or S\textsubscript{2} states \cite{Niedzwiedzki2017, Cong2008} (although the latter is reported to be negligible in Neu--RC--LH1 \cite{Niedzwiedzki2017}).

Returning to our original question on the mechanism of SF in RC--LH1 complexes from \emph{Rba.~sphaeroides}, we conclude that SF in RC--LH1 complexes occurs \textit{via} `heterofission' \cite{Frank1982}: Crt S\textsubscript{2} excitation produces a shared triplet pair state \textsuperscript{3}Crt$\cdot \cdot$\textsuperscript{3}BChl that is delocalised over both molecules, with triplets residing on both the BChl and the Crt. The constituent triplets decay in tandem. The mechanism of the triplet pair decay is likely to be triplet-triplet annihilation. 

\subsection*{Triplet-triplet annihilation populates BChl~\textit{a} Q\textsubscript{y}}

\begin{figure}
\centering
    \includegraphics[width=\textwidth]{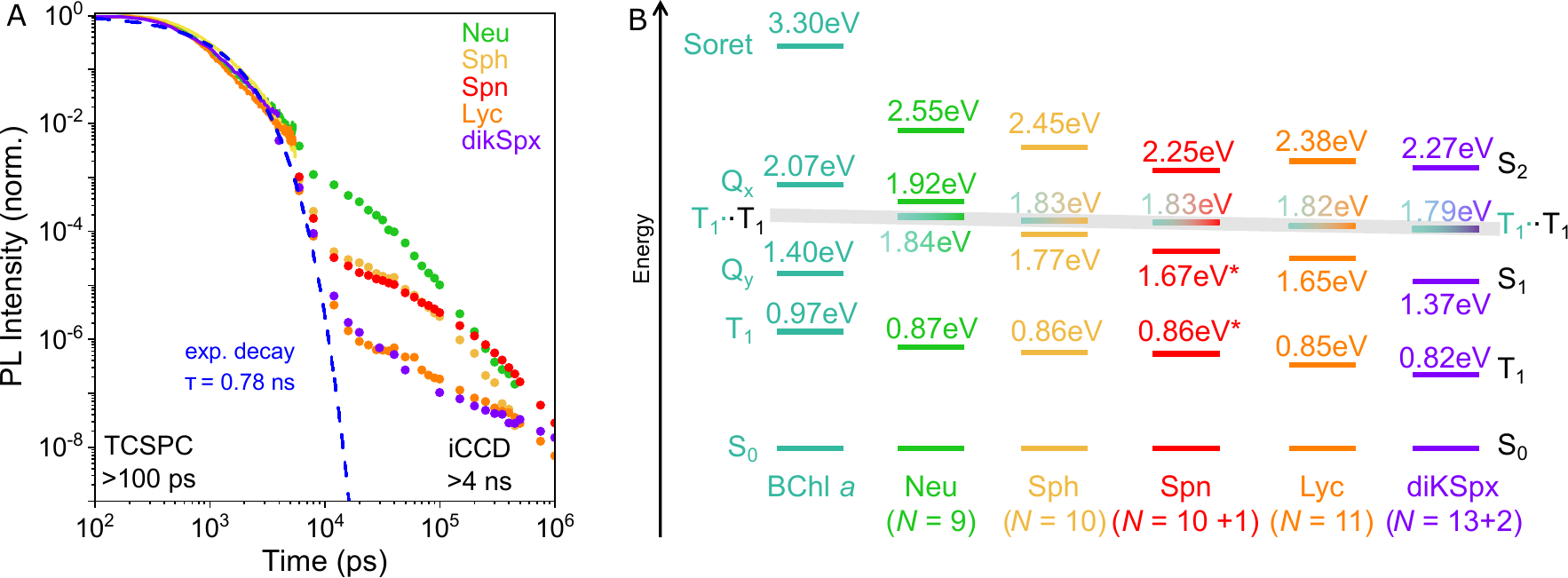}
    \caption{
    \textbf{Fluorescence dynamics and energetics of RC--LH1 (\textit{Rba.~sphaeroides}) as a function of Crt conjugation length}. \textbf{(A)} Fluorescence measured in buffer to avoid scattering effects with Crt excitation (Neu: 500\,nm, others: 520\,nm), detection in the Q\textsubscript{y} emission band (see Fig.~S14 for example spectrum). Solid lines denote TCPSC measurements (normalised at $100$\,ps), markers denote gated intensified CCD measurements (normalised to TCSPC data), dashed line is a monoexponential function for comparison ($\tau = 0.78$\,ns). \textbf{(B)} Approximate energies of BChl~\textit{a} and Crt excited states. Grey shaded line denotes the intermolecular triplet pair energy. Soret, Q-band and Crt S\textsubscript{2} energies from absorption spectra, T\textsubscript{1} and S\textsubscript{1} energies, except Spn, from Refs.~\cite{Fujii1998, Rondonuwu2004, Dilbeck2016, Hartzler2014}. Spn S\textsubscript{1} and T\textsubscript{1} energies are assumed from the conjugation length dependence, see Supplementary Section~8.} \label{fig4}
\end{figure}

Triplet-triplet annihilation (TTA) is the inverse of SF: two triplets fuse to form a single excited state whose spin (singlet or triplet) depends on the relevant spin statistics \cite{Piland2013, Bossanyi2022}. In RC--LH1 complexes, the lowest singlet state is Q\textsubscript{y}, see Fig.~\ref{fig4}~B. TTA, if present, should populate this state, resulting in additional Q\textsubscript{y} emission on the $10-100$\,ns timescale. (Note that we observe no evidence of bimolecular triplet-triplet annihilation in this system, see Fig.~S13, supported by electron paramagnetic resonance (EPR) spectroscopy Ref.~\cite{Bittl2001}, and thus no evidence of triplet migration).

Fig.~\ref{fig4}~A shows transient emission detected in the Q\textsubscript{y} emission band, see example spectra in Fig.~S14. In all complexes, the prompt ($<10$\,ns) BChl~\textit{a} Q\textsubscript{y} emission decays identically, following a monoexponential function (blue dashed line). However, an extra delayed emission component is observed beyond 10\,ns that is dependent on the conjugation length of the bound Crt. Delayed emission from radical pair recombination in the RC \cite{Schenck1982,Rinyu2004} should be present in all complexes, and be independent of Crt conjugation length \cite{Schenck1982}. We therefore attribute the identical delayed emission in the Lyc- and dikSpx-containing samples to radical pair recombination in the RC, confirming our assignment of the MFE (Figs.~\ref{fig2}~(D,E)). On the other hand, the RC--LH1 complexes that demonstrate significant SF (Neu, Sph, Spn) also show an additional delayed emission component that scales with the triplet yield (from Figs.~\ref{fig2}~(F--J)) and occurs on the timescales associated with triplets.

We conclude that the shared triplet pair state, generated by Crt excitation, decays to Q\textsubscript{y}, providing an additional Crt-to-BChl~\textit{a} exciton energy transfer (EET) pathway. 

Interestingly, the relaxed S\textsubscript{1} energy is lower than the triplet pair state in all but the Neu-containing complex, see Fig.~\ref{fig4}~B. Nevertheless, TTA occurs to Q\textsubscript{y} in Sph- and Spn-containing RC--LH1 complexes, and appears not to be quenched by S\textsubscript{1}. One hypothesis to explain this finding is the fact that the S\textsubscript{1} energy depends sensitively on the Crt conformation (e.g.~bond length alternation), which changes as the system relaxes \cite{Taffet2020, Barford2013}. We therefore speculate that TTA to Q\textsubscript{y} is preferred over TTA to S\textsubscript{1}, because S\textsubscript{1} is destabilised compared with Q\textsubscript{y} at the T\textsubscript{1} geometry \cite{Barford2013}. 

\subsection*{Singlet fission as a mechanism for Crt-to-BChl\textit{a} energy transfer}

\begin{figure}
\centering
    \includegraphics[width=0.5\textwidth]{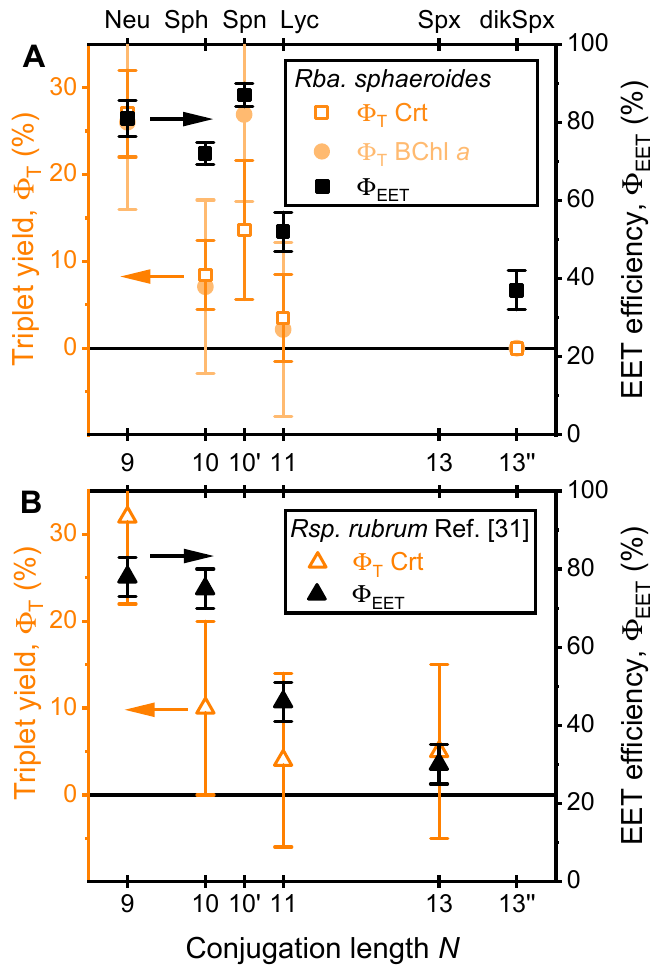}
    \caption{
    \textbf{Triplet yield, $\Phi_T$, and Crt-to-BChl~\textit{a} exciton energy transfer efficiency, $\Phi_{EET}$, in Crt--RC--LH1 complexes as a function of Crt conjugation length, $\textit{N}$} for \textbf{(A)} \emph{Rba.~sphaeroides} and \textbf{(B)} \emph{Rsp.~rubrum}. Data in panel A is from this paper, data in panel B is from Ref.~\cite{Akahane2004}. Details in text and Supplementary Sections~7.1 and 7.2. The general trend is a reduction in both triplet yield and EET efficiency with increasing Crt conjugation length. Note that complexes with high triplet yield show $\Phi_T + \Phi_{EET}>100\%$, indicating that SF must contribute to EET.}\label{fig5}
\end{figure}

We showed above that SF occurs \emph{via} heterofission in Neu--RC--LH1 and the reservoir of SF-generated triplets populates BChl~\textit{a} Q\textsubscript{y} \emph{via} TTA. Thus, SF mediates energy transfer between Crt and BChl~\textit{a}. To determine the importance of this energy transfer pathway, in Fig.~\ref{fig5} we compare the triplet yield ($\Phi_{T}$) after Crt excitation with the Crt-to-BChl~\textit{a} exciton energy transfer efficiency ($\Phi_{EET}$) as a function of Crt conjugation length. Panel A shows results for \emph{Rba.~sphaeroides} and Panel B for \emph{Rsp.~rubrum} \cite{Akahane2004}. 

The triplet yields were obtained from TA spectra in Fig.~\ref{fig2} (\emph{Rba.~sphaeroides}) and Ref.~\cite{Akahane2004} (\emph{Rsp.~rubrum}) using published Crt and BChl~\textit{a} extinction coefficients \cite{Cogdell1983, Borland1987, Slouf2018}, see Supplementary Section~7.2 for details. We note that the absolute Crt triplet yields reported here are generally lower than previous reports \cite{Rademaker1980, Kingma1985, Kingma1985a, Gradinaru2001, Yu2017}, but the trend is robust. 

The Crt-to-BChl~\textit{a} exciton energy transfer efficiency, $\Phi$\textsubscript{EET}, shown in Fig.~\ref{fig5} A, was quantified by comparing 1--transmittance and fluorescence excitation spectra, as described previously \cite{Akahane2004,Chi2015,Niedzwiedzki2017,Noguchi1990,Sutherland2022}, see Supplementary Section~7.1 and Fig.~S16 for details.  Reported values for \emph{Rsp.~rubrum} were taken directly from Ref.~\cite{Akahane2004}. 

The figure indicates that there is a correlation between $\Phi_T$ and $\Phi_{EET}$ and that, apart from Spn--RC--LH1 which has significant ICT character, both the SF yield and EET efficiency drop with increasing conjugation length in both \emph{Rba.~sphaeroides} and \emph{Rsp.~rubrum}. This correlation, together with the fact that $\Phi_T$ and $\Phi_{EET}$  sum to over 100\,\% for the shorter conjugation length Crts, confirms our hypothesis that SF contributes to Crt-to-BChl~\textit{a} exciton energy transfer.

To estimate the relative importance of the SF-mediated energy transfer pathway, we determine approximate yields of all Crt-to-BChl~\textit{a} transfer pathways. These estimates show that for Neu--RC--LH1, roughly $20$\,\% of the overall Crt-to-BChl~\textit{a} energy transfer occurs \emph{via} the SF pathway [9(6)\,\% for Spn--RC--LH1, 18(8)\,\% for Neu--RC--LH1 and 7(4)\,\% for Sph--RC--LH1, see Table~S1]\footnote{Note that these estimates should be treated with caution, hence the large errors, as literature reports on which our calculations are based vary widely. For example, assumed S\textsubscript{2}$\rightarrow$Q\textsubscript{x} EET from Neu to BChl~\textit{a} in RC--LH1, or the energetically similar LH2, varies between negligible \cite{Niedzwiedzki2017} (RC--LH1) through 28\,\% \cite{Cong2008} (LH2) up to 48\,\% \cite{Koyama2004} (LH2). Here, we take a value of $20\pm10$\,\% using scaled overlap factors from Ref.~\cite{Cong2008}, see Supplementary Section~7 for more details.}. Our methods are described in detail in Supplementary Section~7 and yields are tabulated in Table~S1.

\section*{Discussion}

Carotenoids (Crts) play multiple roles in photosynthetic light-harvesting complexes (LHCs). Two key functions are photoprotection \cite{Hashimoto2016, Ruban2007} and light absorption in the spectral gap between the (B)Chl Soret and Q absorption bands \cite{Hashimoto2016}. Photoprotection is facilitated by Crts' ability to rapidly dissipate electronic energy to heat. Due to strong vibronic coupling, Crts can non-radiatively deposit approximately $2-3$\,eV into heat within a mere $10-30$\,ps, two orders of magnitude faster than most fluorescent or acene-based dyes \cite{Chynwat1995, Kakitani2007, Musser2019}. 

Rapid thermal dissipation is useful for photoprotection, but is an issue for harvesting absorbed light. To usefully capture solar energy absorbed by the Crt, the Crt exciton must transfer to a neighbouring BChl~\textit{a} molecule before the energy is dissipated to heat.
We propose that SF-mediated energy transfer offers a way to circumvent rapid dissipation by transiently storing the electronic energy in triplet-pair states, thus enhancing the Crt-to-BChl~\textit{a} energy transfer yield. Weakly exchange coupled triplet states, such as those formed during SF in RC--LH1, are long lived because of spin selection rules that protect them from rapidly decaying to the singlet ground state. While individual triplets are too low in energy to contribute usefully to energy transfer, SF produces \emph{pairs} of (BChl and Crt) triplets, whose combined energy can transfer cooperatively to Q\textsubscript{y}. By circumventing rapid Crt thermalisation and leveraging long-lived spin-protected triplets, we suggest the function of SF is to enhance overall Crt-to-BChl energy transfer efficiency, allowing the organism to usefully harvest more light. 

For SF to be useful in this energy harvesting scheme, triplet pair formation must compete with rapid Crt energy dissipation, i.e. SF should be ultrafast (sub-picosecond). In addition, the generated triplet pair state must not rapidly decay to the ground state. This combination is unusual: ultrafast SF is almost always associated with rapid triplet-pair decay, in the absence of triplet migration \cite{Yong2017, Musser2013, Pun2019, Wang2021, Greissel2023}, due to the strong electronic couplings involved \cite{Smith2013, Musser2019, Taffet2020b}.

Our results indicate that SF occurs directly from S\textsubscript{2}, forming triplet pairs within a few hundred femtoseconds (Fig.~\ref{fig3}~C).  
This direct S\textsubscript{2}-to-triplet-pair formation also explains the conjugation-length-dependent triplet yield (Fig.~\ref{fig5}), as SF must compete with S\textsubscript{2}$\rightarrow$S\textsubscript{1} internal conversion (IC) \cite{Polli2004} and energy transfer to Q\textsubscript{x}, the rates of which are strongly dependent on conjugation length \cite{Polli2004, Cong2008}. 

We also find that despite ultrafast SF, which should result from strong exchange coupling initially \cite{Bossanyi2024, Bayliss2016, Neef2023}, triplet pairs in these complexes are measured to be weakly coupled (Fig.~\ref{fig2}) and long-lived (Fig.~\ref{fig3}). Thus, dynamic localisation, likely driven by heterofission and strong vibronic coupling, decouples the triplets over time, minimising orbital overlap and exchange coupling \cite{Bayliss2016}. This process, akin to relaxation along the combined Crt-BChl potential energy surface, rapidly generates long-lived, weakly coupled triplet pairs, characteristics that are also essential for SF-based technologies \cite{Rao2017, Nagaya2024}. Heterofission thus offers an alternative to triplet migration strategies used in synthetic systems for dynamic triplet-pair decoupling \cite{Yong2017, Musser2013, Pun2019, Wang2021, Greissel2023}.

Recent studies on synthetic SF systems have not only discovered ways of controlling the coupling between triplet pairs \cite{Collins2019, Pun2019, Wang2021, Ha2022, Greissel2023, Ishii2024}, they also emphasise the important interplay between electronic couplings, intermolecular arrangement and dynamic dielectric environment \cite{Kollmar1993,Barford2022, Smith2013, Tamura2015, Taffet2020b, Young2020}. The delicate balance between these parameters can determine whether SF or charge transfer prevails \cite{Young2020}. In the context of photosynthesis, this interplay may explain the contrasting behaviour of RC--LH1 and light-harvesting 2 (LH2) complexes from the same organism \cite{Yu2017}, which share similar Crt and BChl~\textit{a} pigments, but which behave differently upon Crt excitation: RC--LH1 demonstrates SF, but LH2 often favours charge transfer, forming a Crt\textsuperscript{+}$\cdot \cdot$BChl\textsuperscript{--} radical pair \cite{Polivka2004a, Cong2008, Niedzwiedzki2016}.
The charge-transfer process in LH2 is hypothesised to function as an energy quencher \cite{Polivka2004a}. It will be interesting to explore how the detailed electronic/physical structure of the different LHCs enables them to tune between this putative quenching charge-transfer (LH2) and the energy-harvesting SF processes (LH1) described here.

\section*{Conclusions}

In this work, we have demonstrated that the sophisticated structure of purple bacteria light-harvesting complexes (LHCs) enables both ultrafast singlet fission (SF) and the generation of long-lived triplet excitons through a process known as heterofission, involving neighbouring carotenoid (Crt) and bacteriochlorophyll (BChl) pigments. These long-lived triplets act as an energy reservoir, temporarily storing energy to avoid rapid thermal dissipation, enhancing Crt-to-BChl energy transfer by up to $20$\,\%. 

While SF has been predominantly observed in photosynthetic purple bacteria, which often inhabit low-light anaerobic environments where efficient light harvesting is crucial, the widespread presence of Crts and (B)Chls across diverse photosynthetic organisms suggests that similar SF mechanisms may be occurring in other natural systems. Furthermore, the ability of photosynthetic LHCs to dynamically decouple triplet pairs through heterofission, or potentially even to tune between SF and charge transfer by altering pigment-protein structure, offers new strategies for understanding SF and for designing \cite{Krishna2024} new materials for SF-inspired applications \cite{Nagaya2024, Ishii2024}.

\backmatter

\bmhead{Acknowledgements}
We thank E. Taffet, D. I. G. Bennett and W. Barford for insightful discussion. We also thank T. Hayward for help with the electromagnet.

\textbf{Funding:}
This work was supported by the Engineering and Physical Sciences Research Council grant nos. EP/S002103/1 (S.W., G.A.S., C.N.H. and J.C.), EP/T012455/1 (G.A.S., E.C.M,  R.K.V, J.P.P., C.N.H. and J.C.), EP/L022613/1 and EP/R042802/1 (S.W., J.P.P., R.K.V., D.C. and J.C.), and the European Research Council Synergy grant no.~854126 (G.A.S., D.J.K.S., E.C.M. and C.N.H.). C.V. is supported by the Biotechnology and Biological Sciences Research Council (BB/V006630/1). A.H. is a Royal Society University Research Fellow (award number URF/R1/191548). D.J.G. and A.I.T. are supported by the Engineering and Physical Sciences Research Council (EP/S030751/1, EP/V007696/1) and the European Graphene Flagship Project (881603).

\textbf{Author contributions:}
J.C. conceived the study. S.W., G.A.S. and J.C. designed the experiments. G.A.S., E.C.M. and D.J.K.S. grew all the cells and prepared all the protein samples under the supervision of C.N.H. and A.H. S.W. and J.P.P. performed the static and time-resolved spectroscopic measurements. S.W., J.C., J.P.P. and G.A.S. analysed the data. D.J.K.S. assisted with the fluorescence excitation measurements. C.V. performed the TCSPC measurements. D.J.G. and A.I.T. performed and supervised the resonance Raman measurements, respectively. D.C. and J.P.P. assisted with the TA setup. R.K.V. built the NOPA required for MFE measurements. S.W., J.C. and G.A.S. wrote the manuscript and prepared the figures with input from all authors.

\textbf{Competing interests:} The authors declare no competing interests.

\textbf{Data and materials availability:} The datasets underlying all figures in this Research Article and the Supplementary Material are available in the University of Sheffield’s ORDA repository (hosted by figshare, XX).

\bibliography{Main_Text_SF_in_photosynthesis}

\end{document}


\beginsupplement

\section{Materials and Methods} \label{sec:materials}
\subsection{RC--LH1 complexes containing a series of carotenoids} 

\textit{Rba.~sphaeroides} was grown and RC--LH1 complexes were purified according to our method previously described at length \cite{Sutherland2022}. Briefly, for wild-type and mutant stains, a single colony from an M22 agar plate was used to inoculate a 10\,mL universal medical specimen tube, containing M22 liquid medium. This was incubated in the dark at $34\,^\circ$C with agitation (150\,rpm) until visibly turbid. This culture was then scaled up to 80\,mL, also in liquid M22 medium, in 125\,mL Erlenmeyer flasks. Cells were then either cultured by semi-aerobic heterotrophic growth or anaerobic phototrophic growth, as described below.

For semi-aerobic heterotrophic growth, the 80 mL culture was added to 1.6\,L M22 liquid medium in a 2\,L Erlenmeyer flask and incubated at $34\,^\circ$C with agitation (150\,rpm). For anaerobic phototrophic growth, 1\,L M22 medium in a 1.2\,L medical flat was sparged with filtered ultrapure \ce{N2} under aseptic conditions. The 80\,mL culture was added and the flask filled with additional sparged M22 medium. Flasks were then sealed gently with a rubber stopper and parafilm, agitated with a magnetic stirrer and illuminated with white light ($50-500\,\mu$mol photons m$^{-2}$ s$^{-1}$) at room temperature.

Cells were harvested by centrifugation ($4,400 \times$ g, 30\,min, $4\,^\circ$C), resuspended in resuspension buffer (20\,mM HEPES $\mathrm{pH} 8.0$, 5\,mM EDTA), and lysed by twice passing through a French pressure cell (18,000 psi, pre-cooled to $4\,^\circ$C). This solution was centrifuged ($27,000 \times$ g, 15 min, $4\,^\circ$C) and the supernatant was collected. Membranes were obtained by loading the supernatant onto 15/40\,\% (w/v) sucrose step gradients, prepared in the appropriate tubes for a Beckman Ti45 fixed angle rotor, and centrifuged at $85,000 \times$ g, $4\,^\circ$C for $10-16$\,h. Chromatophore membranes were harvested from the interface between the two sucrose solutions, diluted six-fold in resuspension buffer, and centrifuged at $185,000\times$ g for 2\,h. All supernatant was removed and membrane pellets were resuspended and homogenised in 5\,mL resuspension buffer. N-dodecyl-$\beta$-D-maltoside ($\beta$-DDM) was added to a final concentration of 2\,\% (w/v) and membranes were solubilised at $4\,^\circ$C for 1\,h. 

Solutions of resuspension buffer containing 20\,\%, 21.25\,\%, 22.5\,\%, 23.75\,\%, 25\,\%, 50\,\% (w/w) sucrose were prepared in tubes appropriate for a Beckman SW41 Ti rotor. Solubilised membranes were loaded on top of the gradient and centrifuged at $125,000 \times$ g at $4\,^\circ$C for 40\,h. The appropriate band for RC--LH1 dimers (or monomers where appropriate) was collected with a fixed needle.

Protein samples were further purified by anion exchange chromatography using a 5\,mL Q-sepharose HP column (Cytiva) and a linear gradient of buffer A (50\,mM HEPES $\mathrm{pH} 8.0$, 1\,M \ce{NaCl}, 0.03\,\% (w/v) $\beta$-DDM) to buffer B (50\,mM HEPES $\mathrm{pH} 8.0$, 1\,M \ce{NaCl}, 0.03\,\% (w/v) $\beta$-DDM) over 60 column volumes. Appropriate fractions were pooled, concentrated by centrifugal dialysis, and loaded onto a Superdex 200 Increase 10/300 GL size exclusion chromatography column (Cytiva), pre-equilibrated in 50\,mM HEPES $\mathrm{pH} 8.0$, 100\,mM \ce{NaCl}, 0.03\,\% (w/v) $\beta$-DDM. Apart from the carotenoidless RC--LH1 sample, fractions with an A875/A280 ratio $>\,1.4$ were pooled for analysis.

For deposition in trehalose/sucrose films, the RC--LH1 samples were concentrated to give A875 values of $10-20$ (apart from carotenoidless RC--LH1, which was approximately 0.5). 40\,$\mu$L of sample was then mixed with 40 $\mu$L of trehalose/sucrose mix (both 0.5\,M) and 5\,mM sodium ascorbate. Two imaging spacers (Secure Seal diameter 9\,mm, thickness 0.12\,mm; Grace Bio-Labs) were then stacked on a quartz-coated glass substrate (Ossila) and the RC--LH1 trehalose solution was deposited in the centre of the spacer. The substrate was placed in a vacuum chamber (-70\,kPa) with a large excess of calcium sulfate desiccant (Drierite) and dried for a minimum of 3 days. Following drying, a No. 1 coverslip was applied to the imaging spacer to protect the glass from atmospheric rehydration.  

\subsection{Static spectroscopy}
Static absorption spectra were carried out using an Agilent Cary 60 spectrometer.
Static fluorescence excitation spectra were acquired with a Jobin--Yvon Fluorolog--3 spectrofluorometer equipped with a Hamamatsu FL-1030 cooled photomultiplier. The spectra were recorded at $898$~nm with excitation and emission slits set to $3$ and $10$~nm, respectively. To minimise reabsorption, the optical density at $872$~nm for each sample in buffer solution was adjusted to $<0.1$. 
The fluorescence excitation spectra were collected with a short exposure time and multiple accumulations ($\sim 7$\,h in total). 

The resonance Raman spectra were measured with film samples at room temperature using a free-space optical set-up. Briefly, the spectra were recorded \emph{via} an SP750 spectrograph (Princeton Instruments) equipped with a holographic 1800\,g/mm grating and a back-illuminated liquid nitrogen cooled charge-coupled device (Princeton Instruments, PyLon). Excitation at 532\,nm was provided by a single mode diode-pumped solid-state laser (Cobolt, 04-01) with power of 1\,mW focussed in a spot of 2\,$\mu$m diameter on the sample and spectral linewidth of $<1$\,MHz. The spectral resolution of the system was $\sim 0.4\, \mathrm{cm}^{-1}$.

\subsection{Time-resolved absorption (TA) spectroscopy}
The picosecond transient absorption spectra were recorded utilizing a Ti:Sapphire laser system (Spitfire ACE PA-40, Spectra-Physics) delivering pulses with pulse lengths of 40\,fs (FWHM) at 800\,nm and a repetition rate of 10\,kHz. Pump pulses at 500\,nm (and 520\,nm) were generated in an optical parametric amplifier (TOPAS Prime, Light Conversion). Probe pulses were obtained by supercontinuum generation (SCG) in $\ce{CaF2}$ using the 800\,nm Ti:Sapphire fundamental and the fundamental alone. The pump--probe time delay was set using a computer-controlled multi-pass motorised linear delay stage. The polarization of the pump was set to the magic angle (54.7\,$^{\circ}$) to the polarization of the probe. After the sample film, the probe pulses were detected by using a commercial instrument (Helios, Ultrafast Systems) equipped with CMOS detector for the $340-830$\,nm spectral region. 

The nanosecond broadband absorption setup was based on a 90~fs, 1\,kHz Ti:Sapphire amplifier (Solstice, Spectra-Physics). Broadband supercontinuum probe pulses between $450-700$\,nm were generated in a sapphire plate. Pump pulses were provided by an Nd:YVO4 Q-switched laser (Innolas Picolo-AOT) tuned for second-harmonic (532\,nm) output. The pump--probe delay was controlled electronically using a digital delay generator (DG645, Stanford). The polarization of the pump was set to the magic angle (54.7\,$^{\circ}$) with respect to the polarization of the probe. Signal and reference probe pulses created by a beamsplitter were detected spectrally resolved pulse-by-pulse by two CCD cameras (S7030, Hamamatsu) driven and read out at the full laser repetition rate by a custom-built
board from Entwicklungsb\"{u}ro Stresing.

\subsection{Time-resolved photoluminescence (PL) spectroscopy}
The BChl~\textit{a} Q\textsubscript{y} fluorescence lifetime of the RC--LH1 complex was measured on a home-built time-resolved fluorescence microscope. The microscope was equipped with a 485\,nm picosecond diode laser (PicoQuant, PDL 828) as an excitation source. The excitation light was focused by a $100\,\times$ objective (PlaneFluorite, NA $= 1.4$, oil immersion, Olympus). The emitted light was filtered using a 495\,nm dichroic beam-splitter (Semrock) and 900/32\,nm bandpass filter (Semrock) to remove the background excitation light. The microscope was fitted with a spectrometer (150 lines/mm grating, Acton SP2558, Princeton Instruments) and an electron-multiplying charge-coupled device (EMCCD) camera (ProEM 512, Princeton Instruments) for emission spectrum acquisition. A hybrid detector (HPM-100-50, Becker \& Hickl) was used for single photon counting. The modulation of the excitation laser was synchronised with a time-correlated single-photon counting (TCSPC) module (SPC-150, Becker \& Hickl) for the lifetime decay measurement. The repetition rate of the laser was set at 1\,MHz. The excitation laser power was adjusted to produce a fluence of approximately $2\times10^{14}$ photons pulse$^{-1}$ cm$^{-2}$ The instrument response function (IRF) of the set-up was approximately 130\,ps.

The nanosecond-millisecond time-resolved photoluminescence spectra were recorded \emph{via} a time-gated intensified charge-coupled device (iCCD; iStar DH334T-18U-73, Andor), coupled to a Shamrock 303i spectrograph. Excitation pulses at 490 and 520\,nm were generated from a home-built noncollinear optical parametric amplifier (NOPA) pumped by a 1\,kHz regeneratively amplified 800\,nm Ti:Sapphire laser (Solstice, Spectra-Physics).

\subsection{Magnetic field-applied photoluminescence spectroscopy}
The magnetic field effects on the photoluminescence of the RC--LH1 complex were measured by recording the photoluminescence intensity changes at a series of magnetic field strengths ($0-300$\,mT). The film samples were placed between the poles of a magnet (Newport Instrument, Serial no. 8248/II). Excitation pulses at 490 and 520\,nm were created from a home-built NOPA. The photoluminescence was detected at the wavelength $\sim 880$\,nm and the delay time $20-40$\,ns by a spectrograph (Shamrock 303i, Andor) and a time-gated iCCD (iStar DH334T-18U-73, Andor). A 830\,nm longpass filter (FEL830, Thorlabs) was placed before the detector to cut out scattered pump light. The magnetic field effects on the RC--LH1 complex were evaluated as \(\frac{\Delta PL}{PL} = \frac{PL_B - PL_0}{PL_0}\).

\pagebreak

\newpage
\section{Resonance Raman spectra of RC--LH1 complexes indicates identical structural features for all carotenoids} 

According to extensive literature on resonance Raman spectroscopy of carotenoids (e.g.~\cite{Macernis2015, Yu2017, Ruban1995, Ruban2000, Lutz1987}), the weak $\nu_{4}$ bands between $900-980$\,cm\textsuperscript{-1} can be used as a `fingerprint' of molecular conformation \cite{Macernis2015, Yu2017, Ruban1995, Ruban2000, Lutz1987}. This is because the $\nu_{4}$ band is due to \ce{C-H} out-of-plane wagging motions coupled with \ce{C=C} torsional modes which, for planar molecules, are expected to be unconjugated with the electronic transition, and, accordingly, should be vanishingly small in resonance Raman spectroscopy. However, the $\nu_{4}$ band gains intensity when the molecules are distorted around \ce{C-C} bonds \cite{Macernis2015, Yu2017, Ruban1995, Ruban2000, Lutz1987}. Some authors reasonably suggest that the amplitude of the $\nu_{4}$ peak relative to the neighbouring $\nu_{3}$ peak (from \ce{C-CH3} vibrations) can be used as an indication of the twist of the carotenoids in the LH complexes \cite{Papagiannakis2003}.

\begin{figure}[!htbp]%
\centering
\includegraphics[width=0.6\textwidth]{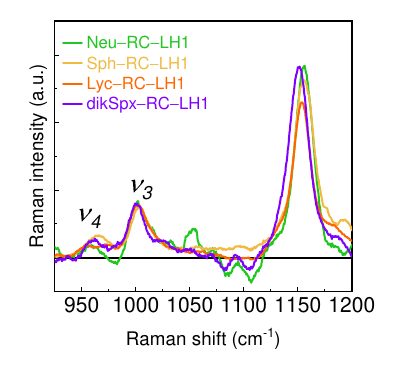}
\caption{Resonance Raman (RR) spectra of RC--LH1 complexes in trehalose films containing neurosporene (Neu), spheroidene (Spe), lycopene (Lyc), and 2,2'-diketo-spirilloxanthin (dikSpx) upon excitation at $\lambda_{\mathrm{ex}} = 532$ nm. The data have been background-subtracted, smoothed using an Adjacent Averaging filter (30 points) and normalised to the $\nu_{3}$ peak at $\sim1000$\,cm\textsuperscript{-1}. Note that owing to the scatter from the films, the background subtraction is not perfect, particularly for Neu--RC--LH1. }\label{RRSpec}
\end{figure}

We used resonance Raman spectroscopy to probe this fingerprint region ($\sim960$\,cm\textsuperscript{-1}) in our samples. See Fig.~\ref{RRSpec}. All four Crts in the RC--LH1 complex measured here show a similar fingerprint region, with the same relative $\nu_{4}$ amplitude compared with $\nu_{3}$, within experimental error, suggesting that the Crts are maintained in a similar, twisted, configuration within the RC--LH1 complexes.

%

\section{Trehalose sugar films stabilise the protein structures but have no impact on their photophysical properties.} \label{sec:trehalose}

As mentioned in the main text, trehalose forms an optically-clear glass film and is used to protect cellular structures under dehydration conditions by maintaining protein tertiary conformation in the absence of water \cite{Kurashov2018}.  The static spectra of the Sph--RC--LH1 complex either encapsulated in trehalose films or dissolved in buffer solutions are similar, indicating that the solid trehalose film does not affect the photophysical properties of RC--LH1 complexes. Note that light scatter is often observed in the $350-600$ nm range as precisely controlling the glass thickness, and hence path length, is not possible using the above method. 



We note that we measure the Q$_\text{y}$ absorption band of RC--LH1 complexes from \textit{Rba.~sphaeroides} to be at 872\,nm, instead of usually reported 875\,nm. To confirm the reproducibility of this static spectral feature we prepared these mutant complexes by two different methods: photosynthetically and semi-aerobically, and both batches show the same 872\,nm Q$_\text{y}$ band. We note that despite the shift, the ratio between the 872\,nm and 275\,nm bands is the same as in previous work: $\sim\,1.6$.

\begin{figure}[!htbp]%
\centering
\includegraphics[width=0.6\textwidth]{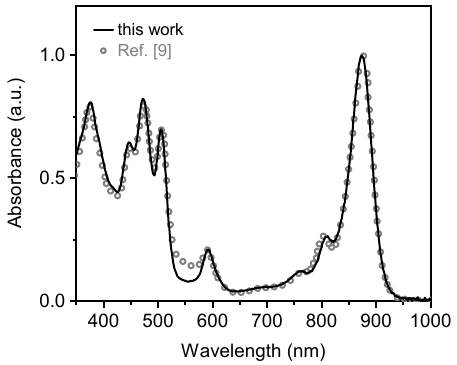}
\caption{Static absorption spectra of the RC--LH1 complex containing spheroidene in a trehalose film (line) and in buffer solution (circles) \cite{Yu2017,Kakitani2007}}\label{trehalose}
\end{figure}

\pagebreak

\section{Transient absorption study of BChl~\textit{a} excited singlet and triplet state} \label{sec:BChl}
\subsection{Difficulties to observe the spectral signatures of BChl~\textit{a} excited triplet state in light-harvesting complex}

\begin{figure}[!htbp]%
\centering
\includegraphics[width=0.5\textwidth]{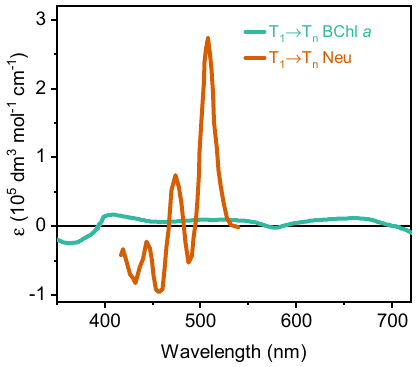}
\caption{Extinction spectra of BChl~\textit{a} T\textsubscript{1}$\rightarrow$T\textsubscript{n} excited state absorption reproduced from Ref.~\cite{Borland1987}, and Neu T\textsubscript{1}$\rightarrow$T\textsubscript{n} excited state absorption reproduced from Ref.~\cite{Cogdell1983}}\label{extinction}
\end{figure}

We found that the Crt signatures dominate the transient absorption spectra of Crt--RC--LH1 complexes (see main text Figs.~2, 3), despite the presence of significant BChl~\textit{a} population. This is well-known \cite{Gradinaru2001,Yu2017}. Empirically, the reason for this is that Crt molecules have relatively large excited state extinction coefficients compared with those of BChl \textit{a}. For example, the Crt T\textsubscript{1}$\rightarrow$T\textsubscript{n} excited triplet state peak absorption extinction coefficient is estimated to be roughly ten times larger ($\sim 2 \times 10 ^5$\,M$^{-1}$\,cm$^{-1}$ \cite{Cogdell1983}) than the BChl~\textit{a} peak triplet absorption coefficient ($\sim 0.2 \times 10 ^5$\,M$^{-1}$\,cm$^{-1}$ \cite{Borland1987}), see Fig.~\ref{extinction}. 
The transient absorption signals of Crt--RC--LH1 in the visible spectral region are therefore dominated by Crt excited state features and the signatures of BChl~\textit{a} are obscured, making assignment difficult. 

\subsection{Spectral similarity of BChl~\textit{a} excited singlet and triplet states} \label{subsec:spectral_similarity}
It is well-known that porphyrin and chlorin molecules show similar singlet S\textsubscript{1}$\rightarrow$S\textsubscript{n} (in this case Q\textsubscript{y}$\rightarrow$S\textsubscript{n}) and triplet T\textsubscript{1}$\rightarrow$T\textsubscript{n} excited state spectral features \cite{Kosumi2018}. This similarity is unusual for $\pi$-conjugated molecules and makes assignment of the spectral features in the congested RC--LH1 transient absorption spectra difficult.
Therefore, here we measured RC--LH1 complexes from a carotenoid-free mutant to obtain reference spectra for Q\textsubscript{y}$\rightarrow$S\textsubscript{n} and  T\textsubscript{1}$\rightarrow$T\textsubscript{n}.

\begin{figure}[!hb]
\centering
\includegraphics[width=\textwidth]{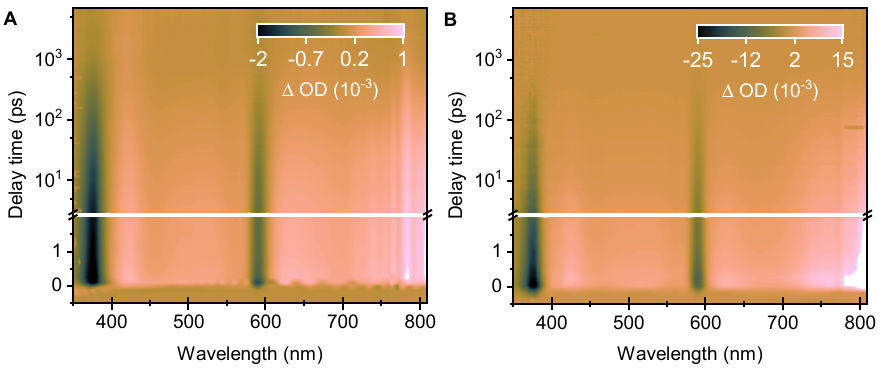}
\caption{Two-dimensional transient absorption maps showing the change in optical density ($\Delta$OD) of the Crt--RC--LH1 complex after excitation at (\textbf{A}) 590\,nm and (\textbf{B}) 870\,nm. The results are displayed using a linear time scale from $-0.5$ to 3\,ps and a logarithmic time scale from $3-7000$\,ps.}\label{2D_Crtless}
\end{figure}

Fig.~\ref{2D_Crtless} shows two-dimensional maps of the transient absorption spectra for a carotenoidless RC--LH1 (Crtless--RC--LH1) complex from \textit{Rba.~sphaeroides} after Q\textsubscript{x} (590\,nm) and Q\textsubscript{y} (860\,nm) excitation. We observe that the BChl~\textit{a} ground state bleach signals at 375\,nm and 590\,nm last within the entire time window (7\,ns). This long-lived BChl~\textit{a} excitation is assumed to be due to its triplet state generated \emph{via} intersystem crossing.

\begin{figure}[!htbp]%
\centering
\includegraphics[width=0.85\textwidth]{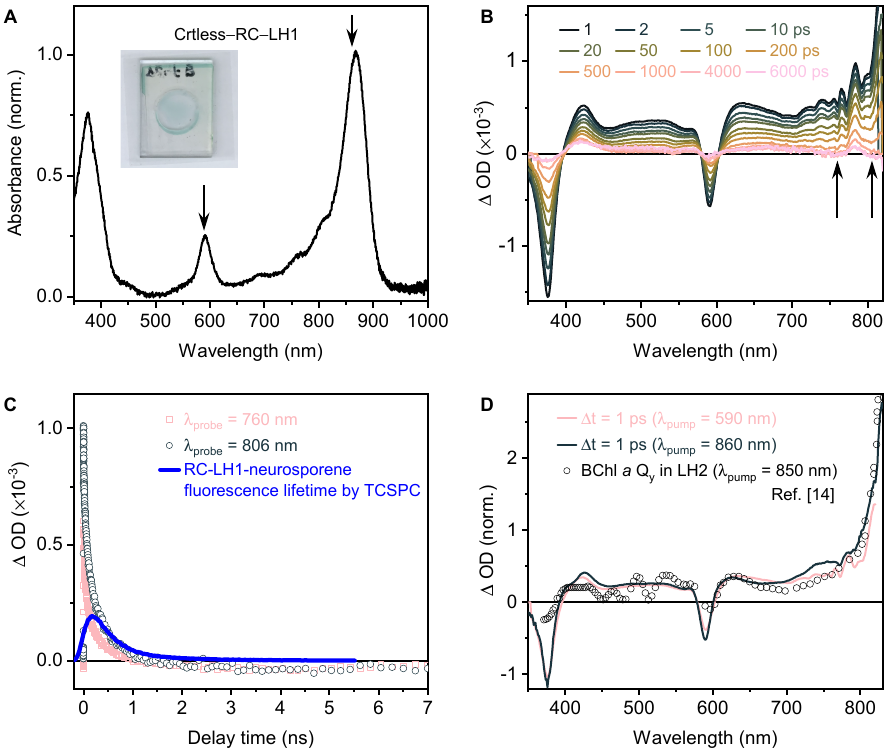}
\caption{(\textbf{A}) Static absorption spectrum of the Crtless--RC--LH1 complex. Excitation wavelengths (590 and 860\,nm) are indicated by vertical arrows. Inset: photograph of the sugar film. (\textbf{B}) Transient absorption spectra of the Crtless--RC--LH1 complex after BChl~\textit{a} Q\textsubscript{x} excitation at 590\,nm for delay times between $1-6000$\,ps. Vertical arrows indicate the selected probe wavelengths (760 and 806\,nm) for plotting the corresponding dynamics in panel (C). (\textbf{C}) Dynamics at selected wavelengths of 760 and 806\,nm decaying to negative after 1\,ns. Also plotted is the fluorescence decay (900\,nm) measured with TCSPC, normalised to 0.7\,ns. Note that TCSPC has a lower temporal resolution which obscures the early-time (sub-200\,ps) decay. (\textbf{D}) BChl~\textit{a} Q\textsubscript{y} spectrum obtained at early time ($\Delta \mathrm{t} =1$\,ps) after both Q\textsubscript{x} (590\,nm) and Q\textsubscript{y} (860\,nm) excitation and the reported BChl~\textit{a} Q\textsubscript{y} excited state spectrum measured in LH2 complex from \textit{Rba.~sphaeroides}, from Ref.~\cite{Kosumi2016}}\label{Crtless}
\end{figure}

During the entire time window in Fig.~\ref{Crtless}B, the spectral features change very little in shape. However, further inspection demonstrates differences in the region between $750-850$\,nm as a function of time. In this spectral region, we observe positive excited state absorption (ESA) at early times and negative ground state bleach (GSB) at late times, see main text Fig.~3C for normalised spectra. The dynamics at selected probe wavelengths 760\,nm and 806\,nm are plotted in Fig.~\ref{Crtless}C showing the $\Delta$OD sign change beyond 1\,ns and the presence of a long-lived (non-decaying) bleach feature at these wavelengths. Importantly, the early-time decay ($< 1$\,ns, ESA signal) matches with the BChl~\textit{a} Q\textsubscript{y} fluorescence lifetime (blue line, Fig.~\ref{Crtless}C). This match shows that the positive excited state absorption (ESA) band between $750-850$\,nm must be associated with Q\textsubscript{y} population. 
Fig.~\ref{Crtless}D shows that this ESA band does not depend on excitation wavelength (when measured at 1\,ps, following internal conversion from Q\textsubscript{x} with 590\,nm excitation), and is similar to the reported Q\textsubscript{y} spectrum from Ref.~\cite{Kosumi2016} measured in LH2 complexes from \textit{Rba.~sphaeroides}. 

We therefore confidently assign the ESA band between $750-850$\,nm to Q\textsubscript{y}$\rightarrow$S\textsubscript{n} absorption. In RC--LH1 Q\textsubscript{y} decays with a $\sim\,1$\,ns time constant. The nature of this decay is discussed further below.

\begin{figure}[!htbp]%
\centering
\includegraphics[width=\textwidth]{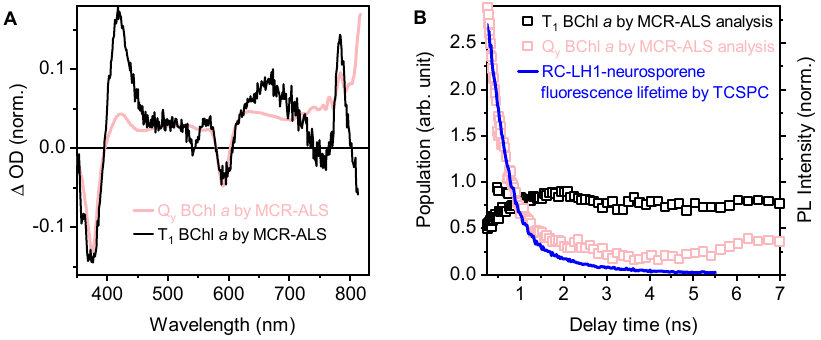}
\caption{Extraction of Q\textsubscript{y} and T\textsubscript{1} components from transient absorption data in Fig.~\ref{Crtless} using MCR--ALS (see text): extracted spectra (\textbf{A}) and extracted dynamics (\textbf{B}).}\label{MCR-ALS-T}
\end{figure}

For a more quantitative understanding of the BChl~\textit{a} excited state spectral behaviour, we performed global analysis using multivariate curve resolution-alternating least squares (MCR--ALS), see Refs.~\cite{Jaumot2005, Jaumot2015} for more details. This MCR--ALS analysis was run on the Crtless--RC--LH1 complex dataset after excitation into the BChl~\textit{a} Q\textsubscript{x} band. We extracted two components with corresponding spectra and dynamics shown in Fig.~\ref{MCR-ALS-T}. We assign the pink component in Fig.~\ref{MCR-ALS-T}A to Q\textsubscript{y} as it closely matches the spectra reported in Fig.~\ref{Crtless} and because its population decay is similar to the measured Q\textsubscript{y} fluorescence (blue curve in panel B).

As the black spectral component in Fig.~\ref{MCR-ALS-T} rises as the Q\textsubscript{y} feature decays, barely decays over $1-7$\,ns and has spectral features of BChl~\textit{a}, we assign it to the triplet T\textsubscript{1} excited state spectrum.

\begin{figure}[!htbp]%
\centering
\includegraphics[width=0.7\textwidth]{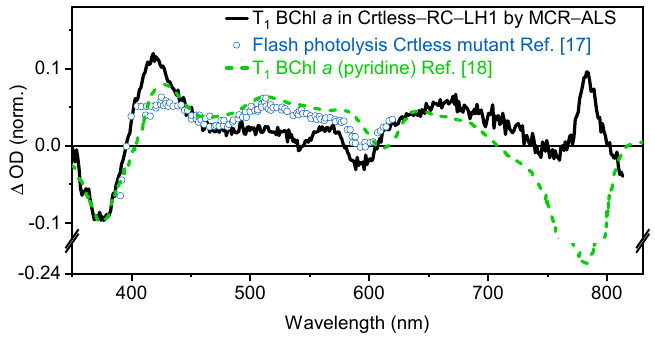}
\caption{Comparison of MCR--ALS-extracted BChl~\textit{a} T\textsubscript{1} spectrum (black) with flash photolysis ($\mu$s) excited state spectrum of a Crtless mutant \textit{Rba.~sphaeroides} chromatophore \cite{Monger1976} (blue circles) and the T\textsubscript{1} spectrum of BChl~\textit{a} in pyridine, measured with flash photolysis \cite{Niedzwiedzki2010}. In RC--LH1, this Q\textsubscript{y} feature is red-shifted compared with the solution.
} \label{MCR-ALS}
\end{figure}

Comparison with reported BChl~\textit{a} T\textsubscript{1} spectra are shown in Fig.~\ref{MCR-ALS}. Although we cannot compare the entire spectral range due to the shift of the bleach from solution (green dashed) to RC--LH1 (our measurement), we find that reported spectra broadly reproduce the visible spectral components.
In addition, absorption-detected microwave resonance (ADMR) spectra of triplets in the reaction centre from \textit{Rba.~sphaeroides} show similar (though shifted) oscillating absorption/bleach in the NIR spectral range \cite{Hartwich1995}.

\begin{figure}[!htbp]%
\centering
\includegraphics[width=0.5\textwidth]{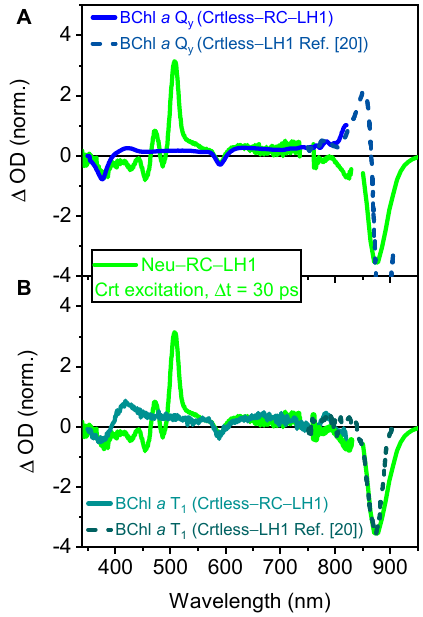}
\caption{(A) Comparison of the LH1-bound BChl~\textit{a} Q\textsubscript{y} spectrum (here, solid) and from Ref.~\cite{Uragami2020} (dashed) together with the Neu--RC--LH1 complex 30\,ps spectrum ($\lambda_{\mathrm{ex}} = 500$\,nm), Crt). (B) Similar comparison with the BChl~\textit{a} T\textsubscript{1} spectrum. 
} \label{fig:QyT1spectra}
\end{figure}

Finally, we note that triplet and Q\textsubscript{y} spectra in Crtless--LH1 complexes have also been published in Ref.~\cite{Uragami2020}, focusing on the Q\textsubscript{y} bleach region. We plot these spectra together with our own, normalised at 785\,nm, in the main text Fig.~3C and here in Fig.~\ref{fig:QyT1spectra}.  These spectra are from Crtless--LH1 complexes from \textit{Rsp.~rubrum}, known to show very similar protein/pigment structure and photophysics compared to LH1 complexes studied here (from \textit{Rba.~sphaeroides}). We find that the spectra from Ref.~\cite{Uragami2020} overlap well with ours between $750-850$\,nm and demonstrate that the Q\textsubscript{y} state shows significantly more GSB intensity between $850-900$\,nm than the BChl~\textit{a} triplet T\textsubscript{1} state. 

In sum, we conclude that the spectra represented in Fig.~3C of the main text (which are almost identical to the MCR--ALS-extracted spectra in Fig.~\ref{MCR-ALS-T}) can be assigned to BChl~\textit{a} Q\textsubscript{y} and T\textsubscript{1} excited state spectra. Q\textsubscript{y} shows a strong absorption between $750-850$\,nm while the T\textsubscript{1} spectrum is dominated by weak ground-state bleach in this region. 

Comparing these reference spectra directly with the 30\,ps Neu--RC--LH1 spectra with Crt excitation ($\lambda_{\mathrm{ex}} = 500$\,nm) in Fig.~\ref{fig:QyT1spectra} suggests that the initial BChl~\textit{a} excitations in Neu--RC--LH1 show more similarity with the BChl~\textit{a} triplet than with the Q\textsubscript{y}.

\newpage
\subsection{Non-linear fluence dependence of BChl~\textit{a} transient absorption spectroscopy} \label{subsec:Crtless BChlexc}

It has been reported that porphyrins have a propensity to demonstrate efficient step-wise two-photon absorption through the S\textsubscript{1} state (the Q state) \cite{Tobita1984}. This has been reported to occur even with relatively low power densities \cite{Tobita1984}, and may need to be controlled for.

\begin{figure}[!htbp]%
\centering
\includegraphics[width=\textwidth]{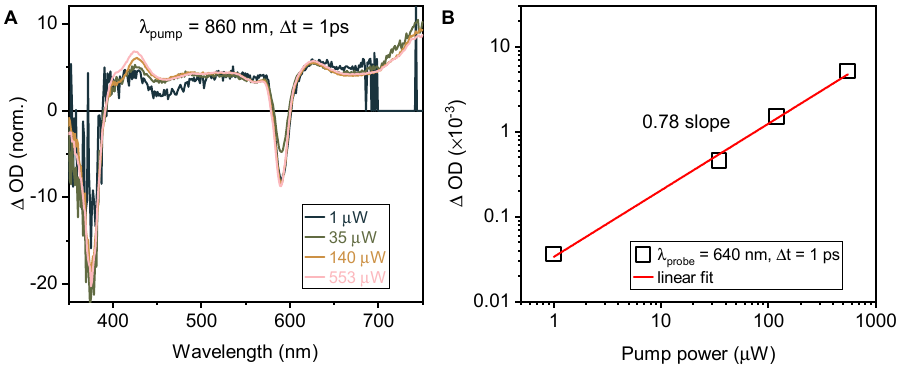}
\caption{(\textbf{A}) Normalised transient absorption spectra of the Crtless--RC--LH1 complex from \textit{Rba.~sphaeroides} for delay time of $1$\,ps after BChl~\textit{a} Q\textsubscript{y} excitation at 860\,nm with pump power from $1-553\,\mu$W (200\,$\mu$m diameter pump spot, 5\,kHz rep.~rate). (\textbf{B}) Dependence of the ESA signal at 640\,nm on the pump pulse power. The red line in the log-log plot is the best fit of the experimental data (squares) and has a slope of 0.78. 
} \label{BChl_Qy}
\end{figure}

Fig.~\ref{BChl_Qy}A shows the normalised transient absorption spectra of the Crtless--RC--LH1 complex recorded at a delay time of $\Delta \mathrm{t} = 1$\,ps after Q\textsubscript{y} excitation at 860\,nm at the power from $1-553$\,$\mu$W. The excitation power dependence of the excited state signal of Q\textsubscript{y} state at the wavelength of 640\,nm and delay time of 1\,ps is shown in Fig.~\ref{BChl_Qy}B.

The figure shows that BChl~\textit{a} in RC--LH1 demonstrates weak non-linear behaviour at fluences typical of our measurements. We observe both a sub-linear increase in ESA intensity with laser fluence (panel B) and a change in the normalised spectrum (panel A), likely associated with two-photon or sequential excitation to form states other than Q\textsubscript{y}. Determining the reason for the spectral shape change at increased fluence is beyond the scope of this work. However, this non-linear behaviour (two photons generate a single excited state) means that triplet yield measurements \cite{Yu2017} which use pulsed lasers should ensure that non-linear effects such as this are controlled for.

\pagebreak
\section{Transient absorption spectroscopy for the RC--LH1 complex after Crt excitation} \label{sec:Car}

\subsection{Global analysis} \label{sec:global analysis}

\begin{figure}[!htbp]%
\centering
\includegraphics[width=0.8\textwidth]{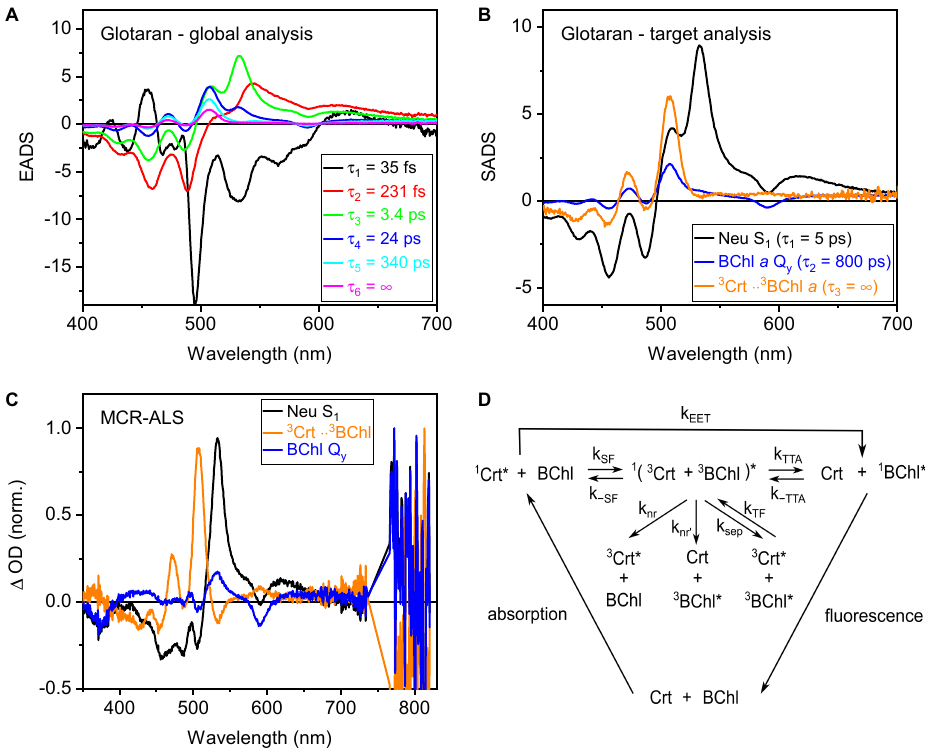}
\caption{(\textbf{A}) Evolution-associated absorption difference spectra (EADS) of photoexcited Neu--RC--LH1 obtained \emph{via} singular value decomposition-based Glotaran global analysis \cite{Snellenburg2012}. The dataset of $400-700$\,nm, $-0.5-7000$\,ps was used for comparison with literature \cite{Niedzwiedzki2017}. (\textbf{B}) Species-associated absorption difference spectra (SADS) obtained by Glotaran target analysis using a kinetic model, see (D). The dataset of $1-7000$\,ps was used.
(\textbf{C}) Species spectra extracted by MCR--ALS global analysis. The dataset of $350-820$\,nm, $1-7000$\,ps was used to simplify the photophysics of excluding of Neu S\textsubscript{2} and BChl~\textit{a} Q\textsubscript{x} states. (\textbf{D}) Simplified kinetic scheme following Crt excitation, rates (k$_\mathrm{i}$) are defined by this scheme}\label{global analysis}
\end{figure}

The molecular dynamics of the RC--LH1 complex containing neurosporene were quantitatively evaluated by a Glotaran global analysis \cite{Snellenburg2012}. An adequate fitting with six components was performed to compare with the previous results on Neu--RC--LH1 from Niedzwiedzki \emph{et al.} \cite{Niedzwiedzki2017}. The global analysis results of evolution-associated absorption difference spectra (EADS) with associated time constants are shown in Fig.~\ref{global analysis}A. The spectra and time constants are in very good agreement with that in Ref.~\cite{Niedzwiedzki2017}. The target analysis unfortunately failed to present the species spectra that we observed in the raw data. For example, the species (Fig.~\ref{global analysis}B) with a time constant of $\sim 5$\,ps is assigned to Neu S\textsubscript{1} but the target-analysis spectrum also contains the contributions from Neu T\textsubscript{1} and BChl~\textit{a} states.

In addition to using Glotaran, which relies on first-order rate target models, we implemented model-free MCR--ALS analysis. We truncated the dataset to $1-7000$\,ps to exclude the initial S\textsubscript{2} and Q\textsubscript{x} states to concentrate on the states we are more concerned with here: S\textsubscript{1}, Q\textsubscript{y}, and (\textsuperscript{3}BChl$\cdot \cdot$\textsuperscript{3}Neu). The MCR--ALS-extracted spectra are displayed in Fig.~\ref{global analysis}C. Unfortunately, we observe clear correlations between spectra. Although the algorithm produced a Neu S\textsubscript{1} spectrum that is similar to reference spectra, correlations were always found to be present at 520\,nm and 780\,nm in the other two spectra (orange and blue). The algorithm cannot therefore successfully separate states beyond S\textsubscript{1}, perhaps because the spectra are too similar or their signals are too weak (compared to the Neu transient signals). We ran the algorithm with different starting conditions, different numbers of states and different data sets, but could find no combination that generated uncorrelated spectra.

\subsection{Carotenoid excited state absorption spectra}
To identify that the carotenoid triplet state is generated, we did the transient absorption experiments on RC--LH1 complexes containing neurosporene (Neu), spheroidene (Sph), lycopene (Lyc), and  2,2'-diketo-spirilloxanthin (dikSpx) and compared the transient spectra with that in the literature of sensitisation experiments on the corresponding carotenoid in organic solvent \cite{Rondonuwu2003}.

\begin{figure}[!htbp]%
\centering
\includegraphics[width=0.6\textwidth]{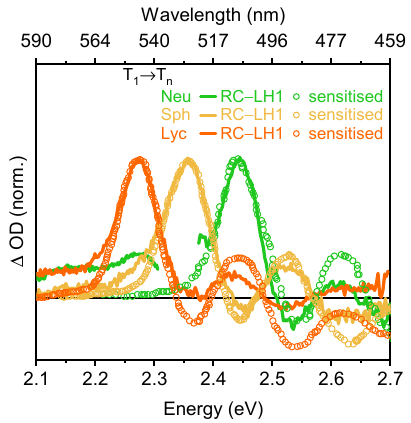}
\caption{Transient absorption spectra of the RC--LH1 complexes containing neurosporene (Neu), spheroidene (Sph), and lycopene (Lyc) (solid, $\Delta \mathrm{t} = 5\,\mu$s) were measured by nanosecond transient absorption spectroscopy. Carotenoid excited triplet state spectra were obtained at 1.8\,$\mu$s after triplet-sensitised excitation (circles) \cite{Rondonuwu2003}. 
}\label{CrtT1}
\end{figure}

The carotenoid T\textsubscript{1} $\rightarrow$ T\textsubscript{n} excited state absorption bands were obtained by nanosecond pump--probe setup after excitation at 532\,nm at delay time of 1\,$\mu$s (Fig.~\ref{CrtT1}), which are in line with the spectra after triplet-sensitised excitation. The peak (0--0 vibration) of carotenoid excited triplet state absorption locates at 507\,nm for Neu, 526\,nm for Sph, 546\,nm for Lyc, and $\sim 625$\,nm for dikSpx. The excited triplet state decays are slow within the displayed time window, while the signal of the S\textsubscript{1} band disappears within $<20$\,ps as seen in the main text (Fig.~2). It should be noted that the apparent initial decay of the carotenoid triplet band is caused in part by the 0--1 vibronic band of the carotenoid $\mathrm{S}_1$ state. 

\begin{figure}[!htbp]%
\centering
\includegraphics[width=0.7\textwidth]{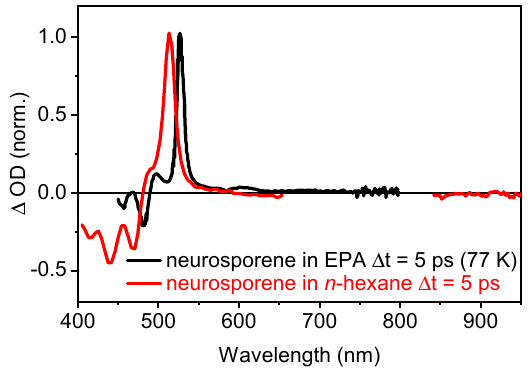}
\caption{Transient absorption spectra of neurosporene at the delay time of $\Delta \mathrm{t} = 5$\,ps. Neurosporene was studied in EPA (diethyl ether/isopentane/ethanol, 5/5/2, v/v/v) at 77\,K (black) \cite{Niedzwiedzki2009} and in \textit{n}-hexane (red) \cite{Zhang2001a}.
}\label{CrtS1}
\end{figure}

Fig.~\ref{CrtS1} \cite{Niedzwiedzki2009,Zhang2001a} shows transient absorption spectra of Neu in organic solvents taken at a delay time of $\Delta \mathrm{t} = 5$\,ps and in the visible and near-infrared region after S\textsubscript{2} excitation. At 5\,ps the spectra exhibit ground state bleach and S\textsubscript{1}$\rightarrow$S\textsubscript{n} excited state absorption.  No obvious transient signals were detected in the near-infrared ($\sim$ 800\,nm) region or at 650\,nm.

The sensitisation and transient absorption experiments on Crts were carried out in organic solvents, such as \textit{n}-hexane and EPA \cite{Rondonuwu2003,Niedzwiedzki2009,Zhang2001a}. Carotenoids in light-harvesting protein which is a highly polarizable environment are in close contact with BChl~\textit{a} molecules. The spectra of carotenoids in RC--LH1 complexes differ by a general wavenumber shift due to the higher polarizability of the protein environment. The effect of the solvent polarizability on the carotenoid $\mathrm{S_1-S_2}$ electronic transitions \cite{Mendes-Pinto2013} explains the shift in carotenoid transient spectra upon binding to the light-harvesting complex. 

\subsection{Long time (ns--ms) transient absorption spectroscopy} \label{subsec:LTTA}
\begin{figure}[!htbp]%
\centering
\includegraphics[width=0.75\textwidth]{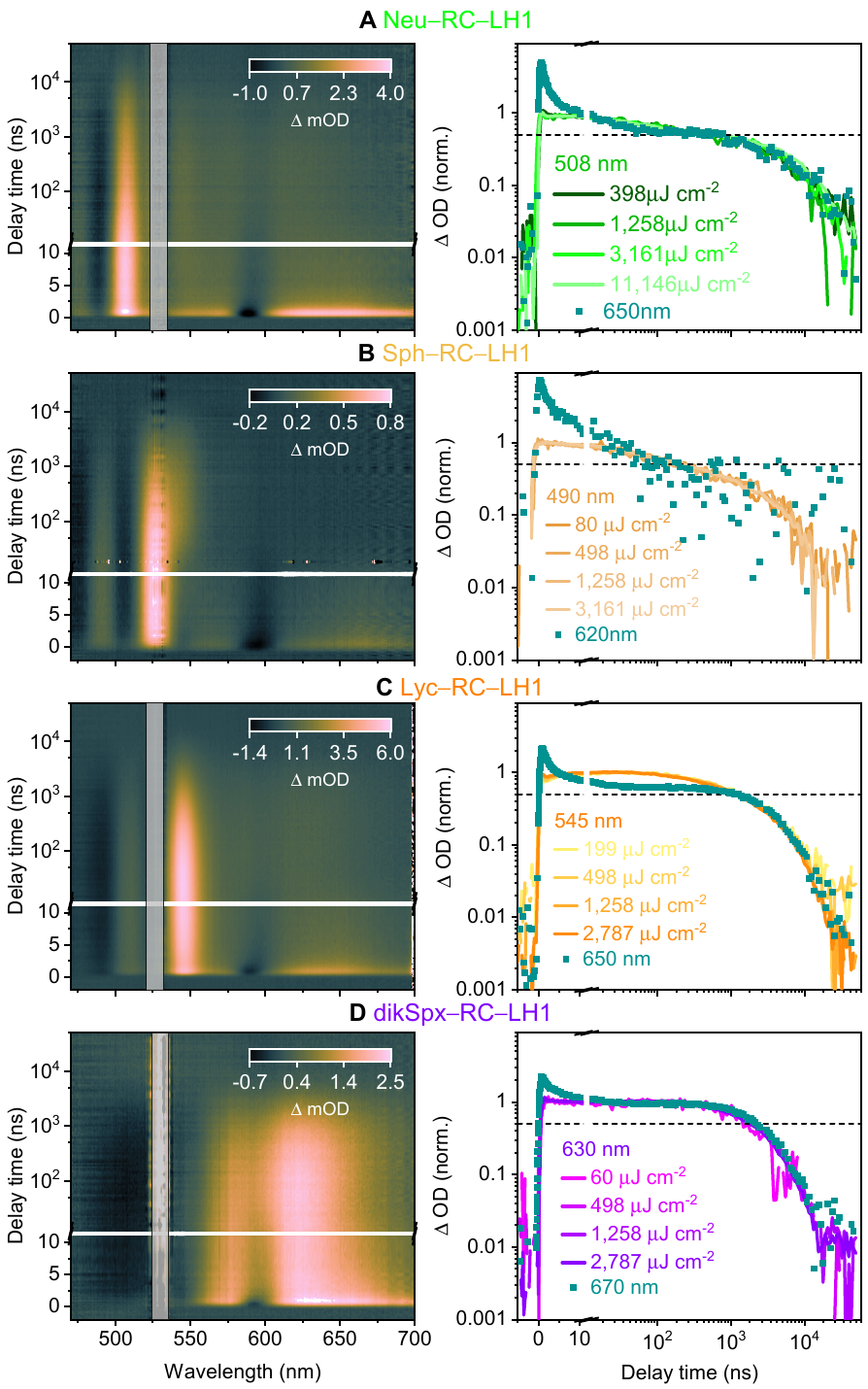}
\caption{ns--ms transient absorption (TA) spectra of RC--LH1 complexes containing (\textbf{A}) neurosporene (Neu), (\textbf{B}) spheroidene (Sph), (\textbf{C}) lycopene (Lyc), and (\textbf{D}) 2,2'-diketo-spirilloxanthin (dikSpx) upon excitation at 532\,nm (500\,ps pulses, $\sim$2\,ns IRF). Left panels show TA maps as a function of time and wavelength. Right panels show normalised kinetic profiles taken at different laser fluences (marked in the legend) within the Crt triplet ESA bands at probe wavelengths 508\,nm (Neu), 490\,nm (Sph), 545\,nm (Lyc), and 630\,nm (dikSpx) and within the much weaker BChl~\textit{a} ESA band (between $620-680$\,nm, teal). Dashed horizontal lines indicate the point at which half the Crt triplet population has decayed. }\label{LTTA}
\end{figure}

The two-dimensional transient absorption maps for RC--LH1 complexes containing neurosporene (Neu), spheroidene (Sph), lycopene (Lyc), and 2,2'-diketo-spirilloxanthin (dikSpx) after excitation at 532\,nm are shown in Fig.~\ref{LTTA} (left panels). The probe spectrum covers the range $470-700$\,nm (supercontinuum generation (SCG) in a sapphire plate). The data were acquired up to $\Delta \mathrm{t} = 50\, \mu$s, using a linear time scale up to 11\,ns and a logarithmic time scale from $11-50000$\,ns. 532\,nm excites both Crt and BChl~\textit{a}, and has weak absorption in the Neu--RC--LH1 sample. 

Each TA map features a prominent long-lived ($\sim\,\mu$s) excited state absorption (ESA) band associated with the Crt triplet as well as weaker signatures of BChl~\textit{a} GSB ($\sim600$\,nm) and ESA across the visible range. As the Crt T\textsubscript{1}$\rightarrow$T\textsubscript{n} transition red-shifts with increasing conjugation length, it becomes difficult to pick out the weak BChl~\textit{a} features from the Crt ones. This is particularly true for dikSpx--RC--LH1.

The time profiles of carotenoid T\textsubscript{1}$\rightarrow$T\textsubscript{n} ESA bands, lines in the right panels of Fig.~\ref{LTTA}, show no excitation density dependence for any of the complexes, indicating that bimolecular processes are not occurring.

The Neu and Sph triplets (lines in panels A, B) start to decay from the first few nanoseconds. They show non-exponential decay across the entire time range with a half-life (indicated by the dashed lines) of $\sim0.3-0.4\,\mu$s.

Triplet dynamics in Lyc--RC--LH1 complexes (panel C, orange/yellow lines, $\lambda_\mathrm{probe} = 545$\,nm), on the other hand, show a delayed \emph{rise} over the first few tens of nanoseconds, most likely due to triplet energy transfer from BChl~\textit{a} triplets (generated by intersystem crossing). In dikSpx--RC--LH1 complexes (panel D, purple lines), a less prominent rise is apparent. In the latter two cases, the Crt triplets decay over slower timescales compared with Neu- and Spe-triplets, with half-lives of $\sim1-3\,\mu$s.

The conjugation length dependence of the triplet lifetime typically follows Jortner's gap-law behaviour and we would therefore expect isolated triplets to decay faster as the conjugation length of the Crt increases \cite{Kakitani2007}, i.e.~opposite to what we observe. Other researchers have noted similar non-gap-law behaviour in LH1 complexes containing short $N<11$ Crts \cite{Kakitani2007}, but were unable to account for it. Here, we suggest that the short triplet lifetime in the Neu- and Sph-containing complexes is due to the predominance of SF-generated triplets that decay \emph{via} triplet-triplet annihilation (TTA). Even in weakly exchange coupled triplet pairs, TTA offers a relatively rapid decay path that is often significantly faster than competing non-radiative decay of the individual triplets \cite{Musser2019}.

If the triplets decay \emph{via} TTA, we should observe similar decay signatures in the BChl~\textit{a} excited state absorption (ESA) band ($600-700$\,nm). Unfortunately, these bands are $\sim 10\times$ weaker than the corresponding Crt triplet ESA bands, making it hard to measure them, particularly as the Crt triplet absorption band red-shifts in the longer Crt complexes (dikSpx- and possibly Lyc--RC--LH1). Below we discuss the BChl~\textit{a} signatures.

In Neu--RC--LH1, panel A, the signal measured in the BChl~\textit{a} ESA band region (e.g.~at 650\,nm; teal markers) is unlikely to be overlapped with the Crt ESA bands \cite{Niedzwiedzki2020} and yet follows the Crt triplet dynamics beyond $\sim15$\,ns, suggesting that BChl~\textit{a} and Crt excitations decay in tandem \emph{via} TTA, as discussed in the main text. Before 15\,ns, the dominant BChl~\textit{a} ESA signature decays with the same time constant as our instrument response $\sim2$\,ns and likely originates from BChl~\textit{a} singlet states that are generated by Crt excitation transfer \emph{via} $\mathrm{S_2\rightarrow Q_x\rightarrow Q_y}$ / $\mathrm{S_2\rightarrow S_1\rightarrow Q_y}$ or directly excited by the 532\,nm pulse. Unfortunately, for these ns--ms experiments we are constrained to use 532\,nm excitation. At this wavelength Neu absorbs only weakly, and some of the initially excited states will originate from direct BChl\textit{a} excitation. These decay within 15\,ns.

In Spn--RC--LH1, panel B, the BChl~\textit{a} ESA has a component that also decays in tandem with the Sph triplet from 20\,ns, suggesting a similar concerted TTA mechanism. However, we note that here the BChl~\textit{a} ESA is weaker than in the Neu--RC--LH1 case, and we observe an additional delayed rise of the Sph triplet in the RC (\emph{cis} isomer with T\textsubscript{1}$\rightarrow$T\textsubscript{n} at $\sim550$\,nm). This signature is not present in the other complexes, possibly due to overlapping features, and warrants further investigation, unfortunately beyond the scope of the current work. 

In Lyc--RC--LH1, panel C, the BChl~\textit{a} signature decays over the first $1\,\mu$s, on the same timescales as the Lyc triplet rise, confirming triplet energy transfer from BChl~\textit{a} to Lyc. Beyond $1\,\mu$s both signals decay together. The reason for this parallel decay is difficult to interpret, but may be due to spectral overlap between the Lyc triplet and BChl~\textit{a} ESA bands. 

The dikSpx triplet ESA overlaps so strongly with the BChl~\textit{a} ESA that it is impossible to pick apart the individual components.

\newpage
\section{PL spectra}

\begin{figure}[ht]%
\centering
\includegraphics[width=0.9\textwidth]{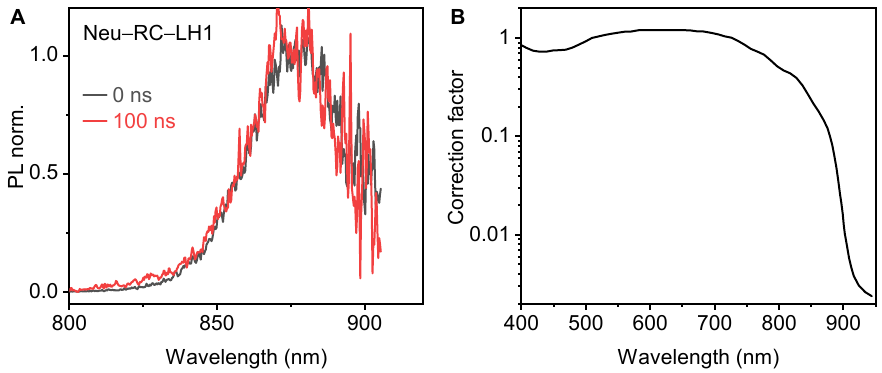}
\caption{(\textbf{A}) Prompt (black) and delayed (red) emission spectrum of the Neu--RC--LH1 complex at delay time $\Delta \mathrm{t} = 0$\,ns and $\Delta \mathrm{t} = 100$\,ns, respectively, after 500\,nm excitation. The BChl~\textit{a} $\mathrm{Q_y}$ emission energy transferred from Crt singlet states and shared (T$\cdot \cdot$T) state appear spectrally unchanged.
(\textbf{B}) The spectral sensitivity of the
iCCD.}\label{PLspec}
\end{figure}

The delayed emission at 100\,ns in Fig.~\ref{PLspec}A is ascribed to the BChl~\textit{a} $\mathrm{Q_y}$ formed by triplet-triplet annihilation. The delayed emission spectrum of BChl~\textit{a} $\mathrm{Q_y}$ after excitation of Crt at 500\,nm matches its fluorescence spectrum after excitation of BChl~\textit{a} at 800\,nm.

The image intensifier of our iCCD employs a Gen III VIH ($360-910$\,nm) photocathode. The photocathode and grating largely determine the spectral sensitivity of the detector.
The BChl~\textit{a} $\mathrm{Q_y}$ emission in the near-infrared region is substantially affected by the spectral sensitivity of the gated iCCD spectrometer. The emission spectra in Fig.~\ref{PLspec}A were corrected for the spectral sensitivity of the iCCD, but as the sensitivity in this region drops of significantly, the spectral shape should be treated with caution.

\newpage
\section{Estimation Crt-to-BChl~\textit{a} exciton energy transfer efficiency} \label{sec:EET-yield}

\begin{figure}[!htbp]%
\centering
\includegraphics[width=0.95\textwidth]{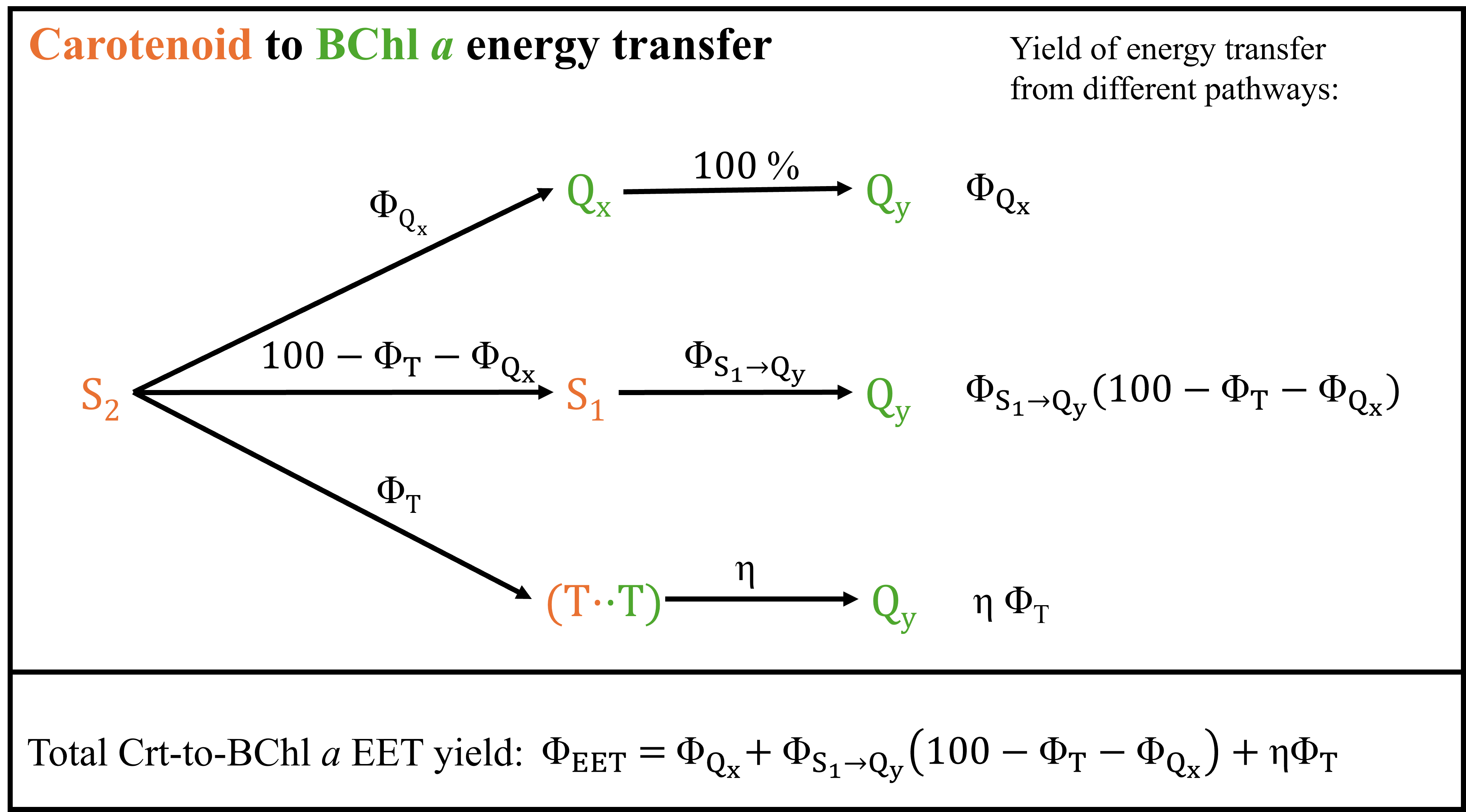}
\caption{Scheme showing the different Crt-to-BChl~\textit{a} exciton energy transfer pathways and relative yields. Here, we assume no losses to the ground state directly from S\textsubscript{2} and no losses during Q\textsubscript{x}$\rightarrow$Q\textsubscript{y} internal conversion. Non-radiative losses to the ground state are assumed to occur during S\textsubscript{1} or (T$\cdot \cdot$T) decay.}\label{EET-scheme}
\end{figure}

Crt-to-BChl~\textit{a} EET occurs \emph{via} three main pathways. Assuming SF occurs directly from S\textsubscript{2}, the scheme in Fig.~\ref{EET-scheme} and associated equation

\begin{equation}
    \mathrm{\Phi_{EET}=\Phi_{Q_x}+\Phi_{S_1\rightarrow Q_y}(100 - \Phi_{T} - \Phi_{Q_x}) + \eta \Phi_{T}}
    \label{eq:EET}
\end{equation}
are used here to quantify the exciton energy transfer efficiency (EET).

To quantify the relative importance of the SF channel \emph{via} (T$\cdot \cdot$T), $\eta\,\Phi\textsubscript{T}$, we must determine the relative yields of the other pathways and the total Crt-to-BChl~\textit{a} EET efficiency. 

Below we determine: (1) $\Phi\textsubscript{EET}$ from fluorescence excitation and absorptance spectroscopy; (2) $\Phi\textsubscript{T}$ from transient absorption spectroscopy; (3) $\mathrm{\Phi_{S_1\rightarrow Q_y}}$ by comparing the lifetime of S\textsubscript{1} in solution and in RC--LH1 complexes; (4) $\mathrm{\Phi_{Q_x}}$ and $\eta$ are difficult to determine. We therefore estimate their values by plotting $\mathrm{\Phi_{Q_x}}$ as a function of $\eta$ and comparing this to estimates of $\mathrm{\Phi_{Q_x}}$ calculated using S\textsubscript{2} and Q\textsubscript{x} spectral overlap from Cong \emph{et al.,} \cite{Cong2008}, normalised to measured $\mathrm{\Phi_{Q_x}}$ for rhodopin glucoside \cite{Krueger1998, Macpherson2001}. The results of these estimates are shown in Table \ref{tbl:EET estimation}, reproduced in the main text.

\begin{table}[!ht]
\centering
\caption{\textbf{Crt-to-BChl Exciton Energy Transfer (EET) efficiencies and associated parameters}. S\textsubscript{1} lifetimes in LH1 complex, $\tau_{\mathrm{S_{1}}}^{\mathrm{LH1}}$ (see Fig.~2) and solvent ($\tau_{\mathrm{S_{1}}}^{\mathrm{Sol}}$) from Refs.~\cite{Cong2008, Zhang2000, Slouf2018} are used to estimate EET efficiency from S\textsubscript{1}$\rightarrow$Q\textsubscript{y} using $
 \Phi_{S_{1}\rightarrow Q_y}=\left( 1-\tau_{\mathrm{S_{1}}}^{\mathrm{LH1}}/\tau_{\mathrm{S_{1}}}^{\mathrm{Sol}}\right)\times100$, as described previously \cite{Cong2008, Zhang2000}. The triplet yield $\Phi$\textsubscript{T} was calculated using data from the transient absorption spectra in Fig.~2 and published extinction coefficients \cite{Cogdell1983}. The overall Crt-to-BChl~\textit{a} EET efficiency, $\Phi_{EET}$, was determined by comparing 1-transmittance and fluorescence excitation spectra in Fig.~\ref{PLE}, as described previously \cite{Akahane2004,Chi2015,Niedzwiedzki2017,Noguchi1990,Sutherland2022}. The S\textsubscript{2}$\rightarrow$Q\textsubscript{x} EET efficiency, $\Phi_{Q_x}$  are best-guess estimates obtained by scaling spectral overlap factors reported in Ref.~\cite{Cong2008} with the experimental value for rhodopin glucoside ($\textit{N}=11$) \cite{Macpherson2001}, see Fig.~\ref{EET_S2-Qx}. These values are used to estimate the efficiency of energy transfer \emph{via} the triplet pairs, $\eta \Phi_T$, see Fig.~\ref{EET_from_SF}. Errors are given in parentheses, estimated from fitting errors, 2$\sigma$ spread or from variability in the literature.}
\label{tbl:EET estimation}
\begin{tabular}{lcccccccc} \\ \cmidrule{1-9}
  & \multicolumn{3}{c}{S$_1$ lifetime (ps)} & \multicolumn{2}{c}{Triplet yield (\%)} & \multicolumn{3}{c}{EET efficiency (\%)} \\ 
\multicolumn{6}{c}{}  & $\mathrm{S_2 \rightarrow Q_x}$ & $\mathrm{S_1 \rightarrow Q_y}$ & $S_2 \rightarrow (\mathrm{T\cdot\cdot T})$ \\ 
 \multicolumn{6}{l}{}  & $\rightarrow Q_y$ &  & $\rightarrow Q_y$ \\
  \cmidrule{1-9} 
 & $\tau_{\mathrm{S_{1}}}^{\mathrm{LH1}}$ & $\tau_{\mathrm{S_{1}}}^{\mathrm{sol}}$  & Ref.& $\Phi_T$ & $\Phi_{EET}$ & $\Phi_{Q_x}$ & $\Phi_{S_1 \rightarrow Q_y}$ & $\eta \Phi_T$ \\
 \cmidrule{1-9} 
Neu & 4.6 (0.4)& 24 & \cite{Cong2008} & 27 (5) & 81 (4.6) & 21 (10) & 81 (2) & 18 (8) \\
Sph & 5.3 (0.6) & 9.3 & \cite{Zhang2000} & 8 (4) &  72 (2.8) & 44 (10) & 43 (6) & 7 (4) \\
Spn & 2.2 (0.4) & 6 & \cite{Cong2008} & 14 (4) & 87 (3) & 64 (10) & 63 (7) & 9 (6) \\
Lyc & 4.0 (0.5) & 4.7 & \cite{Zhang2000} & 3 (5) & 52 (5) & 51 (10) & 15 (10) &  \\
dikSpx & 1.7 (0.4) & 1.5 & \cite{Slouf2018} &  & 37 (5) &  &   &  \\ \cmidrule{1-9}
\label{tbl:EET}
\end{tabular}
\end{table}

\begin{figure}[!htbp]%
\centering
\includegraphics[width=0.95\textwidth]{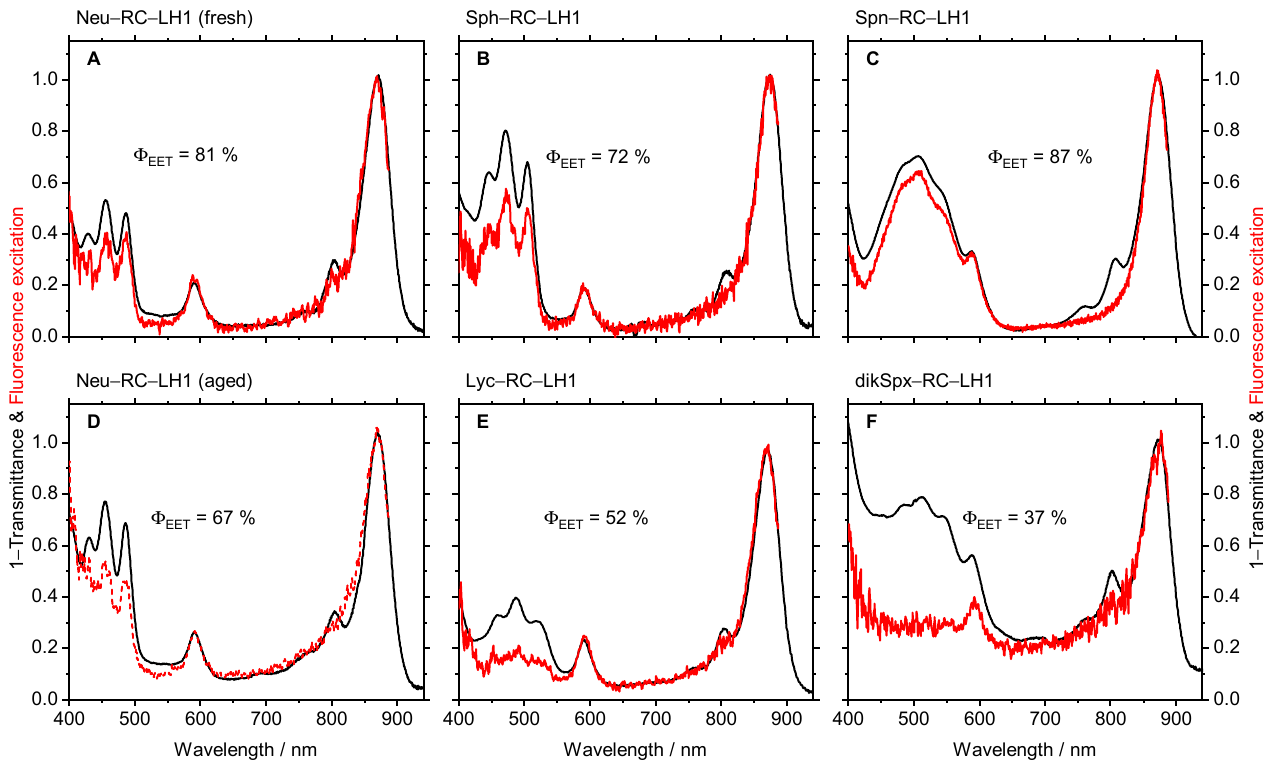}
\caption{1--transmittance (black line) and fluorescence excitation (red line) spectra of RC--LH1 complexes containing Neu (\textbf{A}), Sph (\textbf{B}), Spn (\textbf{C}), Lyc (\textbf{E}), and dikSpx (\textbf{F}). The overall exciton energy transfer efficiency ($\Phi_{\text{EET}}$) is displayed in each panel. We note that these spectra were measured within hours of purification and found that $\Phi$\textsubscript{EET} values obtained on older Neu--RC--LH1 samples had dropped to 67\,\% (\textbf{D}).}\label{PLE}
\end{figure}

\subsection{Total Crt-to-BChl~\textit{a} EET yield ($\Phi$\textsubscript{EET}) determination} \label{sec:EET-overall}

The exciton energy transfer yield $\Phi$\textsubscript{EET} was quantified by comparing 1--transmittance and fluorescence excitation spectra (see Fig.~\ref{PLE}), as described previously \cite{Akahane2004,Chi2015,Niedzwiedzki2017,Noguchi1990,Sutherland2022}. Details of static spectroscopy measurements are provided in section~\ref{sec:materials}.

These spectra are plotted in Fig.~\ref{PLE}. We determined the ratio of the peak intensities at the 0--0 carotenoid  S\textsubscript{0}$\rightarrow$S\textsubscript{2} transition where the contribution from BChl~\textit{a} is minimal to that in the absorptance spectrum. The fluorescence excitation spectra in Fig.~\ref{PLE} were corrected for the excitation lamp spectrum and response spectrum of the detector and then normalised at the BChl~\textit{a} $\mathrm{Q_y}$ band.

The overall Crt EET efficiency listed in Table~\ref{tbl:EET estimation} is the mean value, obtained by averaging 10\,nm around the Crt 0--0 vibronic peak. The error is given by $2\sigma$ (standard deviation). 

The yields are also shown in Fig.~5 in the main text for both \emph{Rba.~sphaeroides} and \emph{Rsp.~rubrum} (the latter data were taken directly from Ref.~\cite{Akahane2004}).

We note that there is a spread in the measured values of $\Phi$\textsubscript{EET} both in the literature \cite{Noguchi1990,Niedzwiedzki2017} and here. For example EET from Neu to BChl~\textit{a} in the RC--LH1 complex from \textit{Rba.~sphaeroides} range from $\Phi_{\mathrm{EET}}=67-81$\,\% (here, Fig.~\ref{PLE}) to $72$\,\% \cite{Noguchi1990} and $55$\,\% \cite{Niedzwiedzki2017}. We have found that aged samples show lower $\Phi$\textsubscript{EET}, Fig.~\ref{PLE}, and therefore the values in Table \ref{tbl:EET estimation} and Fig.~5 in the main text are from fresh samples (measured within hours of purification). 

The trend in Fig.~5 is consistent with literature \cite{Akahane2004,Chi2015}: shorter conjugation length Crts in light-harvesting complexes demonstrate increased EET efficiency. Indeed, except for Spn--RC--LH1, both \textit{Rba.~sphaeroides} and \textit{Rsp.~rubrum} show a decrease in $\Phi$\textsubscript{EET} as the conjugation length of the incorporated Crt increases.  

\subsection{Triplet yield}\label{sec:triplet-yield}

The triplet yield $\Phi$\textsubscript{T} was calculated using data from the transient absorption spectra in Fig.~2 in the main text and similar data extracted from Ref.~\cite{Akahane2004}. The resulting yields are reported in Table~\ref{tbl:triplet yield} and Fig.~5 in the main text. To calculate the yield, concentrations of the triplet state of Crts (Neu, Sph, Spn, and Lyc) and BChl~\textit{a} were directly calculated from TA spectroscopy at pump--probe delay time at $\Delta \mathrm{t} = 100$\,ps and $\Delta \mathrm{t} = 2$\,ns, respectively, by using the Lambert-Beer law. The concentration of the carotenoid S\textsubscript{2} state in each case was estimated by the size of the initial ground state bleach amplitude based on the molar extinction coefficients. The extinction coefficients of carotenoid T\textsubscript{1} and S\textsubscript{2} and BChl~\textit{a} T\textsubscript{1} are listed in Table~\ref{tbl:triplet yield}. To reduce errors, we have used Crt extinction coefficients published in the same work \cite{Cogdell1983} and used the same method to determine the yield in each case. Therefore, although the absolute yield of Crt or BChl~\textit{a} triplets may not be comparable with other reports (our Crt triplet yields are generally lower \cite{Kingma1985,Kingma1985a,Rademaker1980, Gradinaru2001, Yu2017}), the trend with conjugation length is robust. The triplet yields are collected in Table~\ref{tbl:triplet yield} and shown in Fig.~5 in the main text.


\begin{sidewaystable*}[!htbp]
\centering
\caption{Extinction coefficients ($\epsilon$) of Crt T\textsubscript{1} and S\textsubscript{2} and BChl~\textit{a} T\textsubscript{1}, transient signal intensities of excited state absorption (ESA) of Crt T\textsubscript{1} ($\Delta \mathrm{t} = 100$\,ps) and BChl~\textit{a} T\textsubscript{1} ($\Delta \mathrm{t} = 2$\,ns) and initial ground state bleach (GSB) of Crt S\textsubscript{2} for RC--LH1 complexes after Crt S\textsubscript{2} excitation by TA spectroscopy, estimated triplet quantum yields ($\Phi$\textsubscript{T}) of Crt and BChl~\textit{a} generated \emph{via} singlet fission.}
\vspace{7mm}
\label{tbl:triplet yield}
\begin{tabularx}{5\textheight}{lcccccccc}
RC--LH1 complex &
  \begin{tabular}[c]{@{}c@{}}$\epsilon^a$ \\ Crt T$_1$\\ (M$^{-1}$ cm$^{-1}$)\end{tabular} &
  \begin{tabular}[c]{@{}c@{}}$\epsilon^b$ \\ Crt S$_2$\\ (M$^{-1}$ cm$^{-1}$)\end{tabular} &
  \begin{tabular}[c]{@{}c@{}}$\epsilon^c$ \\ BChl~\textit{a} T$_1$\\ (M$^{-1}$cm$^{-1}$)\end{tabular} &
  \begin{tabular}[c]{@{}c@{}}ESA$^{\ddag}$ \\ Crt T$_1$\\ ($\Delta$OD)\end{tabular} &
  \begin{tabular}[c]{@{}c@{}}GSB$^{\S}$ \\ Crt S$_2$\\ ($\Delta$OD)\end{tabular} &
  \begin{tabular}[c]{@{}c@{}}ESA$^{\ddag}$ \\ BChl~\textit{a} T$_1$\\ ($\Delta$OD)\end{tabular} &
  \begin{tabular}[c]{@{}c@{}}$^{\Phi}$ \\ Crt T$_1$\\ (\%)\end{tabular} &
  \begin{tabular}[c]{@{}c@{}}$^{\Phi}$ \\ BChl~\textit{a} T$_1$\\ (\%)\end{tabular} \\
Neu & 2.74E5 & 1.594E5 & 1.036E4 & 2.36E-3 & -5.10E-3 & 8.60E-5 & 27.0 & 26.0 \\
Sph  & 3.09E5 & 1.736E5 & 1.036E4 & 5.68E-4 & -3.80E-3 & 1.60E-5 & 8.4 & 7.1 \\
Spn  & 6.06E4 & 1.220E5 & 1.036E4 & 2.37E-4 & -3.50E-3 & 8.00E-5 & 13.6 & 26.9 \\
Lyc  & 4.95E5 & 1.815E5 & 1.036E4 & 7.67E-5 & -8.12E-4 & 1.00E-6 & 3.5 & 2.2 
\end{tabularx}
\footnotetext{$^a$ from Ref.~\cite{Cogdell1983}, $^b$ from Ref.~\cite{Sashima2000}, $^c$ from Ref.~\cite{Borland1987}, $^{\ddag}$ ESA is for excited state absorption, $^{\S}$ GSB is for ground state bleach.}
\end{sidewaystable*}

\subsection{Estimated EET along the S\textsubscript{1} pathway, $\mathrm{\Phi_{S_1\rightarrow Q_y}}$}\label{sec:S1-EET}

 The S\textsubscript{1}$\rightarrow$Q\textsubscript{y} EET efficiency was estimated by comparing the measured S\textsubscript{1} lifetime from Fig.~2 (bottom panels in the main text) with those reported in solution, as described in literature using the following equation \cite{Cong2008, Zhang2000},
 \begin{equation} 
 \Phi_{S_1 \rightarrow Q_y }=\left( 1-\frac{\tau_{\mathrm{S_{1}}}^{\mathrm{LH1}}}{\tau_{\mathrm{S_{1}}}^{\mathrm{Sol}}}\right)\times100,
 \end{equation}
 yielding Neu: 81\,\%, Sph: 43\,\%, Spn: 63\,\%, Lyc: 15\,\%. These are reported in Table \ref{tbl:EET estimation}.

    \subsection{Estimated EET directly from S\textsubscript{2} to Q\textsubscript{x}, $\mathrm{\Phi_{Q_x}}$ }\label{sec:S2-EET}

\begin{figure}[!ht]%
\centering
\includegraphics[width=0.5\textwidth]{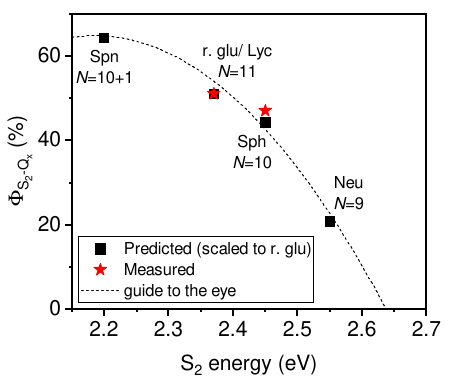}
\caption{Best-guess estimates of S\textsubscript{2}$\rightarrow$Q\textsubscript{x} energy transfer efficiencies (black squares) obtained by scaling spectral overlap factors reported in Ref.~\cite{Cong2008} by the broadly accepted experimental value for rhodopin glucoside ($\textit{N}=11$) \cite{Macpherson2001}. Red stars are measured values for comparison, rhodopin glucoside from Ref.~\cite{Macpherson2001}, Sph from Ref.~\cite{Ricci1996}.}\label{EET_S2-Qx}
\end{figure}

The S\textsubscript{2}$\rightarrow$Q\textsubscript{x} EET efficiency is difficult to estimate, due to the widely different, often model-dependent \cite{Frank2009energy}, values reported in the literature for the same complex. Here, we use best-guess estimates by scaling spectral overlap factors reported in Ref.~\cite{Cong2008} with a broadly accepted experimental value for rhodopin glucoside ($\textit{N}=11$) \cite{Macpherson2001}, see Fig.~\ref{EET_S2-Qx} and Table \ref{tbl:EET estimation}. These estimates avoid problems associated with model-dependent values \cite{Frank2009energy}, but they should nevertheless be treated with caution. The measured \cite{Ricci1996} and predicted values for Sph match within a few percentage points, but further ultrafast measurements are needed for Neu, Spn and other Crts. We suggest a 10\,\% error in these values for the further analysis below.

 \subsection{Estimated $\eta$, the fraction of (T$\cdot \cdot$T) states that form Q\textsubscript{y}}\label{sec:EET-SF}

Finally, using equation \ref{eq:EET}, and the values described above, we estimate EET \emph{via} (T$\cdot \cdot$T).
\begin{figure}[!ht]%
\centering
\includegraphics[width=\textwidth]{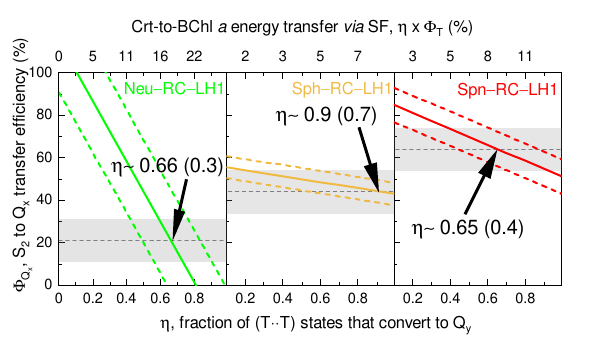}
\caption{Parametric plots relating the fraction of (T$\cdot \cdot$T) states that convert to Q\textsubscript{y}, $\eta$ (x-axis), to the efficiency of S\textsubscript{2} to Q\textsubscript{x} transfer, $\mathrm{\Phi_{Q_x}}$ (y-axis) for given values of all other variables. Left: Neu--RC--LH1, middle: Sph--RC--LH1 and right: Spn--RC--LH1 (right). The coloured lines represent the function from Eq~\ref{eq:EET}: $\mathrm{\Phi_{Q_x}=(-\Phi_{EET}+\Phi_{S_{1}\rightarrow Q_{y}}(100-\Phi_{T})+\eta \times \Phi_{T})/(\Phi_{S_1\rightarrow Q_y}-1)}$ using values from table \ref{tbl:EET estimation} (dashed coloured lines are upper and lower estimates based on uncertainty in $\Phi_{EET}$). The grey dashed lines show estimates of $\mathrm{\Phi_{Q_x}}$ from the previous section (shading indicates uncertainty). Overlap of the two lines provides estimated values of $\mathrm{\Phi_{Q_x}}$ and $\eta$.}\label{EET_from_SF}
\end{figure}
Figure \ref{EET_from_SF} shows a plot of values of $\mathrm{\Phi_{Q_x}}$ as a function of $\eta$ using the values in Table \ref{tbl:EET estimation} and a reformulation of equation \ref{eq:EET}.

For Neu--RC--LH1, $\mathrm{\Phi_{Q_x}}$ and $\eta$ are strongly correlated. Using the estimate from spectral overlap factor $\mathrm{\Phi_{Q_x}= 20\pm10\,\%}$ from Fig.~\ref{EET_S2-Qx} (shown as the grey dashed line in Fig.~\ref{EET_from_SF}) suggests $\eta\approx 0.66\pm 0.3$ and the Crt-to-BChl~\textit{a} EET contribution from SF is $\mathrm{\eta \Phi_{T}\approx 18\pm 8\,\%}$.

For Sph--RC--LH1, $\mathrm{\Phi_{Q_x}}$ is more constrained, but $\eta$ varies from 0 to 1, even when taking into account estimates of $\mathrm{\Phi_{Q_x}}$. Our best-guess is therefore $\eta=0.9\pm0.7$ or $\mathrm{\eta \Phi_{T}\approx7 \pm 3\,\%}$. 

For Spn--RC--LH1, we take $\eta\approx0.65\pm 0.4$ and $\mathrm{\eta \Phi_{T}\approx 9\pm 6\,\%}$.

It is surprising that $\eta$ is estimated to be larger than 0.6 (although a value of 0.6 is within error in all cases). The spin statistical limit for incoherent weakly coupled triplet pairs recombining to form a singlet state \cite{Bossanyi2022} has a maximum value of 0.6. We speculate that $\eta >0.6$ may be possible because nanosecond TTA occurs while (T$\cdot \cdot$T) is coherent, and therefore $\eta=1$ for some time, increasing the time-averaged limit.

\newpage

\section{Discussion of Spn S\textsubscript{1} energy} \label{sec:spheroidenone}

Crt S\textsubscript{1} energies have been reported using different experimental methods. The generally accepted Spn S\textsubscript{1} energy is 1.61\,eV extracted from transient absorption spectroscopy \cite{Cong2008, Zigmantas2004}, but this energy seems inconsistent with our findings in the main text (Fig.~4B). We, therefore, re-examine the estimated Spn S\textsubscript{1} energy.

The generally accepted neurosporene and spheroidene S\textsubscript{1} energies were determined by using fluorescence spectroscopy \cite{Fujii1998}. When we use the reported linear regression of S\textsubscript{1} energies with respect to the conjugation lengths ($\mathrm{E(S_1)} = 220946\times \frac{1}{2\textit{N} + 1}+3681$) we estimate the Spn S\textsubscript{1} energy to be 1.70\,eV ($\textit{N}=10.5$). The S\textsubscript{1} energies of Crt molecules with longer conjugation lengths were also predicted by using S\textsubscript{1} lifetimes and the energy gap-law \cite{Frank1997,Chynwat1995}. Using this energy gap-law we estimate the Spn S\textsubscript{1} energy to be 1.70\,eV ($\tau=6$\,ps \cite{Cong2008}). 

As described above, the Crt S\textsubscript{1} energies were also estimated using transient absorption spectroscopy \cite{Cong2008}, resulting in a Spn S\textsubscript{1} energy of 1.63\,eV. Zigmantas \emph{et al.} used the same technique and reported the energy to be about 1.61\,eV \cite{Zigmantas2004}.

\begin{table}[ht]
\centering
\caption{Crt S\textsubscript{1} energies estimated using different reported methods.}
\label{tbl:S1 energy}
\begin{tabular}{ccc}
\hline
method &
  TA spectroscopy \cite{Cong2008,Zigmantas2004} &
  \begin{tabular}[c]{@{}c@{}}fluorescence spectroscopy \cite{Fujii1998}\\ energy gap law with S\textsubscript{1} lifetimes \cite{Frank1997,Chynwat1995}\end{tabular} \\ \hline
\multicolumn{1}{c}{Crt} &
  \multicolumn{2}{c}{S\textsubscript{1} energy (cm\textsuperscript{-1})} \\
    \cmidrule{2-3}     
Neu     & 14170 & 15300 \\
Sph     & 13160 & 14202 \\
Spn     & 12800 & 13724 \\
Lyc     & --    & 13394 \\
rhodopin glucoside & 12400 & --    \\
Spx     & --    & 11926 \\ \hline
\end{tabular}
\end{table}

With the comparison of the two datasets (Table~\ref{tbl:S1 energy}), it is noted that the transient absorption (TA) results always show a blue shift compared with the fluorescence or energetic gap-law values. Indeed, we find that the TA-obtained value matches the fluorescence 0--1 peak instead of the 0--0 peak. This inconsistency is interesting but we leave further discussion of it to later work and note only that the Spn S\textsubscript{1} energy follows the conjugation length trend of other Crts, despite the presence of an ICT state.

\newpage

\bibliography{Main_Text_SF_in_photosynthesis}